\documentclass[twocolumn]{aastex631}
\usepackage{import}
\usepackage{amsmath}
\usepackage{amssymb}

\newcommand\aastex{AAS\TeX}

\newcommand{\ropt}{$R_{0.5\mu\rm{m}}$}

\newcommand{\rmass}{$R_{\rm{M_\star}}$}
\newcommand{\rmassd}{$R_{\rm{M_\star,3D}}$}
\usepackage{enumerate}

\received{January 1, 2999}
\revised{January 2, 2999}
\accepted{January3, 2999}
\submitjournal{ApJ}

\shorttitle{\aastex\ Stellar Half-Mass Radii}
\shortauthors{van der Wel et al.}

\begin{document}

\title{Stellar Half-Mass Radii of $0.5<z<2.3$ Galaxies: Comparison with JWST/NIRCam Half-Light Radii}

\author[0000-0002-5027-0135]{Arjen van der Wel}
\affiliation{Sterrenkundig Observatorium, Universiteit Gent, Krijgslaan 281 S9, 9000 Gent, Belgium}

\author[0000-0003-2373-0404]{Marco Martorano}
\affiliation{Sterrenkundig Observatorium, Universiteit Gent, Krijgslaan 281 S9, 9000 Gent, Belgium}

\author[0000-0002-1857-2088]{Boris H\"au\ss ler}
\affiliation{European Southern Observatory, Alonso de Cordova 3107, Casilla 19001, Santiago, Chile}

\author[0000-0001-5294-8002]{Kalina~V.~Nedkova} 
\affiliation{Department of Physics and Astronomy, Johns Hopkins University, Baltimore, MD 21218, USA}

\author[0000-0001-8367-6265]{Tim B. Miller}
\affiliation{Department of Astronomy, Yale University, 52 Hillhouse Avenue, New Haven, CT 06511, USA}

\author[0000-0003-2680-005X]{Gabriel B. Brammer}
\affiliation{Cosmic Dawn Center (DAWN)}
\affiliation{Niels Bohr Institute, University of Copenhagen, Lyngbyvej 2, 2100 Copenhagen, Denmark}

\author[0000-0003-4546-7731]{Glenn van de Ven}
\affiliation{Department of Astrophysics, University of Vienna, T\"urkenschanzstrasse 17, 1180 Vienna, Austria}

\author[0000-0001-6755-1315]{Joel Leja}
\affiliation{Department of Astronomy \& Astrophysics, The Pennsylvania State University, University Park, PA 16802, USA}
\affiliation{Institute for Computational \& Data Sciences, The Pennsylvania State University, University Park, PA 16802, USA}
\affiliation{Institute for Gravitation and the Cosmos, The Pennsylvania State University, University Park, PA 16802, USA}

\author[0000-0001-5063-8254]{Rachel S. Bezanson}
\affiliation{Department of Physics and Astronomy and PITT PACC, University of Pittsburgh, Pittsburgh, PA 15260, USA}

\author[0000-0002-9330-9108]{Adam Muzzin}
\affiliation{Department of Physics and Astronomy, York University, 4700 Keele Street, Toronto, Ontario, ON MJ3 1P3, Canada}

\author[0000-0001-9002-3502]{Danilo Marchesini}
\affiliation{Department of Physics and Astronomy, Tufts University, Medford, MA}

\author[0000-0002-2380-9801]{Anna de Graaff}
\affiliation{Max-Planck-Institut f\"ur Astronomie, K\"onigstuhl 17, D-69117, Heidelberg, Germany}

\author[0000-0002-7613-9872]{Mariska Kriek}
\affiliation{Leiden Observatory, Leiden University, P.O. Box 9513, 2300 RA, Leiden, The Netherlands}

\author[0000-0002-5564-9873]{Eric F. Bell}
\affiliation{Department of Astronomy, University of Michigan, 1085 South University Avenue, Ann Arbor, MI 48109-1107, USA}

\author[0000-0002-8871-3026]{Marijn Franx}
\affiliation{Leiden Observatory, Leiden University, P.O. Box 9513, 2300 RA, Leiden, The Netherlands}


\correspondingauthor{Arjen van der Wel}
\email{arjen.vanderwel@ugent.be}

\begin{abstract}

We use CEERS JWST/NIRCam imaging to measure rest-frame near-IR light profiles of $>$500 $M_\star>10^{10}~M_\odot$ galaxies in the redshift range $0.5<z<2.3$. We compare the resulting rest-frame 1.5-2$\mu$m half-light radii ($R_{\rm{NIR}}$) with stellar half-mass radii (\rmass) derived with multi-color light profiles from CANDELS HST imaging. In general agreement with previous work, we find that $R_{\rm{NIR}}$ and \rmass~are  up to 40\%~smaller than the rest-frame optical half-light radius $R_{\rm{opt}}$.
The agreement between $R_{\rm{NIR}}$ and \rmass~is excellent, with negligible systematic offset ($<$0.03 dex) up to $z=2$ for quiescent galaxies and up to $z=1.5$ for star-forming galaxies. 
 We also deproject the profiles to estimate \rmassd, the radius of a sphere containing 50\% of the stellar mass. 
We present the $R-M_\star$ distribution of galaxies at $0.5<z<1.5$, comparing $R_{\rm{opt}}$, \rmass~and \rmassd. 
The slope is significantly flatter for \rmass~and \rmassd~ compared to $R_{\rm{opt}}$, mostly due to downward shifts in size for massive star-forming galaxies, while  \rmass~and \rmassd~do not show markedly different trends.
Finally, we show rapid size evolution ($R\propto (1+z)^{-1.7\pm0.1}$) for massive ($M_\star>10^{11}~M_\odot$) quiescent galaxies between $z=0.5$ and $z=2.3$, again comparing $R_{\rm{opt}}$, \rmass~and \rmassd. 
We conclude that the main tenets of the size evolution narrative established over the past 20 years, based on rest-frame optical light profile analysis, still hold in the era of JWST/NIRCam observations in the rest-frame near-IR.
\end{abstract}


\section{Introduction}
Projected light profiles of galaxies are widely used as proxies for their 3-dimensional stellar mass profiles, both at low and high redshifts. Under this assumption, great progress has been made in our understanding of the structure of galaxies and their assembly history. At the same time, we have known for decades that galaxies show color gradients \citep{sandage72, peletier90, de-jong96}, implying that, given the correlation between color and stellar mass-to-light ratio \citep[$M_\star/L$][]{bell01}, $M_\star/L$ varies with radius, and generally peaks in the center for massive galaxies. The gradients arise due to a combination of radial variations in attenuation and stellar population properties (age, abundances, IMF). For early-type galaxies the color gradient is generally understood to be due to a radial variation in metallicity \citep[e.g.,][]{li18}, while for star-forming galaxies the stellar population gradients, in both age and metallicity, are significant but generally mild \citep[e.g.,][]{sanchez-blazquez14}, and centrally concentrated attenuation plays a dominant role, especially at higher redshifts as is now being revealed by JWST \citep{miller22}.

In order to interpret observations of light profiles in the context of theoretical models or simulations, these $M_\star/L$ gradients must be taken into account as their effect on the half- mass radius as inferred from observations can be very substantial
(a factor $\sim2$), even at near-infrared wavelengths \citep[e.g.,][]{dutton11}.  In addition, projected light or mass profiles can be difficult to interpret: a direct comparison with simulations requires a deprojection in three dimensions \citep[e.g.,][]{prugniel97, baes11, cappellari13, van-de-ven21}, or the creation of mock observations by projecting simulated galaxies \citep{de-graaff22}.

The interpretation of the redshift evolution of galaxy light profiles usually relies on the assumption that the $M_\star/L$ gradient does not (strongly) evolve, as most studies ignore color and $M_\star/L$ gradients. Several authors used observed color gradients in higher-redshift galaxies, first reported by \citet{hinkley01} and \citet{mcgrath08}, to address the impact of $M_\star/L$ gradients on galaxy size estimates \citep{mcgrath08, guo11, wuyts12, szomoru13, mosleh17, suess19a, suess19b, mosleh20, miller23}. Generally, the results point to the existence of qualitatively similar color gradients at low and high redshift, but even relatively small changes can strongly affect the interpretation of the observed size evolution of galaxies \citep{suess19b, suess20, miller23}.

Likewise, the evolution of galaxy geometry (the intrinsic, three-dimensional shape) is often overlooked.  \citet{chang13a, van-der-wel14a, zhang19} show that geometry strongly evolves with redshift, which implies that the interpretation of projected light profiles must change with redshift, even if its impact has not yet been analyzed.

With the arrival of JWST we can access for the first time the rest-frame near-IR light profiles of intermediate redshift (here, $0.5<z<2.3$) galaxies that should provide a more direct proxy of the stellar mass profile since attenuation becomes negligible in most cases and variations in $M_\star/L$ as a function of age and metallicity are less strong.  In this paper (Sec.~2) we use the first batch of NIRCam imaging from the CEERS program \citep{finkelstein23} that covers $\approx 4$\%~of the full CANDELS dataset to test the robustness of stellar half-mass radii estimates based on resolved color profiles from HST imaging \citep{koekemoer11, grogin11}. Early work by \citet{suess22} already demonstrated that the rest-frame near-IR sizes from JWST/NIRCam are somewhat smaller than the rest-frame optical sizes as measured from HST/WFC3, supporting the previous results that stellar mass-weighted profiles are more compact than light-weighted profiles. Sec.~\ref{sec:deproject} describes the methodology to convert projected sizes into 3D sizes. In Sec.~3 we present size-mass distributions at for the different size proxies (rest-frame optical, mass-weighted, deprojected) and the average size evolution for massive quiescent galaxies. In Sec.~4 we summarize the results.

We assume a flat $\Lambda$CDM cosmology with $H_0=70$~km s Mpc$^{-1}$ and $\Omega_{\rm{m}}=0.3$, and the \citet{chabrier03} stellar initial mass function.

\section{Data and Methodology}
\subsection{A New Approach for Estimating Stellar Half-Mass Radii}
\label{sec:methods}

A variety of methods has been developed to convert light distributions into stellar mass maps, which can be divided along two axes. First, some methods create 2D mass maps \citep[e.g.,][]{abraham99, zibetti09, meidt12, wuyts12, meidt14, tacchella15, abdurrouf18, suess19a, mosleh20, abdurrouf23}, which have the advantage of retaining all spatial information, while other methods create symmetrized (1D) profiles \citep[e.g.,][]{szomoru13, fang13, tacchella15, suess19a, miller23}, which have the advantage that they are more easily corrected for the PSF and, relevant to the topic at hand, more easily compared with standard methods to measure galaxy sizes.

Second, the spatially resolved photometric information can be converted into $M_\star/L$ information by SED fitting \citep[e.g.,][]{suess19a, mosleh20}, or by the the application of color-$M_\star/L$ relations devised for integrated galaxy light \citep{bell01, van-der-wel05, meidt14} and applied to spatially resolved light distributions \citep[e.g.,][]{zibetti09, meidt14, fang13, szomoru13, tacchella15, suess19a, miller23}. The former has the advantage that all available information is used, but (rest-frame) near-IR photometry is required to assign unbiased $M_\star/L$ values to dusty regions \citep[e.g,][]{zibetti09, meidt14}. The latter has the advantage that color-$M_\star/L$ relations can leverage the knowledge of $M_\star$ obtained from broad-band SED fitting across a wide wavelength range, including the near-IR.
With any method, we should always keep in mind the uncertainties related to choices made to assign a `true' stellar mass, that is, uncertainties in the stellar population synthesis models and the implementation of absorption and scattering by dust -- a discussion of these issues is beyond the scope of this paper.

Since, in our case, we do not have spatially resolved rest-frame near-IR photometric information and the goal is to construct stellar half-mass radii for comparison with light-weighted radii, we choose to analyze 1D profiles and apply newly developed color-$M_\star/L$ relations. Our method consists of the following steps. First, we use $M_\star$ estimates from SED fits over the full available wavelength range (UV-to-mid-IR) as described in Sec.~\ref{sec:sed} to construct a (redshift-dependent) relationship between $M_\star/L$ and multiple HST colors in the observed frame (Sec.~\ref{sec:colorml}). Second, assuming that the same relationship holds within galaxies, we convert HST light profiles (Sec.~\ref{sec:profiles}) into stellar mass profiles via the multi-color-$M_\star/L$~relation (Sec.~\ref{sec:conversion}).

The advantages of this method are multiple. Long-wavelength information is leveraged (via the SED-based $M_\star$ estimates) in a redshift-dependent manner; that is, any evolution in the relationship between color and $M_\star/L$, which is significant \citep{li22}, is automatically included. Furthermore, no conversion from observed to rest-frame colors is required, which removes template-related uncertainties. Finally, rather than a single color we use the shape of the SED for which spatially resolved information is available to estimate $M_\star/L$. At each step we take care to formally propagate the uncertainties, resulting in robust uncertainties on the inferred stellar half-mass radii.

\subsection{Multiwavelength Photometry and SED Fitting}
\label{sec:sed}
The 3D-HST/CANDELS photometric catalog provided by \citet{skelton14} was used by \citet{leja20} to estimate stellar masses, star-formation rates and other physical parameters with the Prospector-$\alpha$~ model \citep{leja17, johnson21}. The model uses the \citet{conroy09} stellar population FSPS, a non-parametric star-formation history, a two-component dust model, and optionally indcludes an enshrouded AGN. 

The Leja et al.~catalog serves as the basis of our work and contains 63413 galaxies in the redshift range $0.5<z<3.0$. This is a stellar mass-complete sample, where the completeness limit increases from $\log(M_\star/M_{\odot})\approx 8.7$ at $z=0.5$ to $\log(M_\star/M_{\odot})\approx 10.1$ at $z=3.0$. We define galaxies as quiescent when their 100 Myr-averaged star-formation rates are 0.8 dex or more below the star-forming sequence as defined by \citet{leja22}.

\subsection{Derivation of Color-$M/L$ Relations}
\label{sec:colorml}
Our novel approach to derive mass profiles gradients rests on the assumption that the color-$M/L$ relation within individual galaxies is identical to the relation among galaxies. We create a (multi-)color-$M/L$ relation based on the full SED fitting results described above, for a set of colors for which we have spatially resolved information from HST. As such we leverage photometry with a much broader dynamic range in wavelength (UV to mid-IR) to infer $M/L$ estimates based on a more limited range for which spatially resolved profiles are available (0.6 - 1.6 micron in the observed frame).

In order to capture the effects of cosmological redshift and evolution of stellar populations simultaneously, we fit a relation of the following form:

\begin{equation}
\begin{split}
    \log{(M_\star/F_{160})} = a_0 + a_1\log{(1+z)} + 
    a_2\log{(F_{606}/F_{160})} \\ 
    + a_3\log{(F_{814}/F_{160})} + a_4\log{(F_{125}/F_{160})}
\end{split}
\end{equation}

\noindent where $M_\star$ is the stellar mass estimate from the full SED fit, $z$ is the redshift, and the F values are the total flux densities from the Skelton catalog in units of AB$=25$ magnitude in the respective HST filters (F606W, F814W, F125W and F160W). The fit minimizes $\chi^2$, which is dominated by the uncertainties in $M_\star$ rather than the photometric data.

\begin{table}
\tabletypesize{\scriptsize}
\begin{center}
  \caption{ Coefficients for Eq.~1}
 \begin{tabular}{|l|c|c|c|c|c|}
\hline
 & $a_0$ & $a_1$ & $a_2$ & $a_3$ & $a_4$ \\
\hline
SF & 7.652  &  2.879   &  0.130     &  0.575   &  0.562  \\
Q  & 7.857  &  2.959   &  0.168     &  0.204   &  0.536 \\
\hline
\end{tabular}
\tablecomments{Coefficients from Eq.~1 that describe the fitted relationship between HST flux density ratios (colors) and $M_\star/F_{160}$ across the redshift range $0.5<z<2.3$ and stellar masses $M_\star>10^{10}~M_{\odot}$.}
\label{tab:color_ml}
\end{center}
\end{table}

\begin{figure}[!h]
\epsscale{1.15}
\plotone{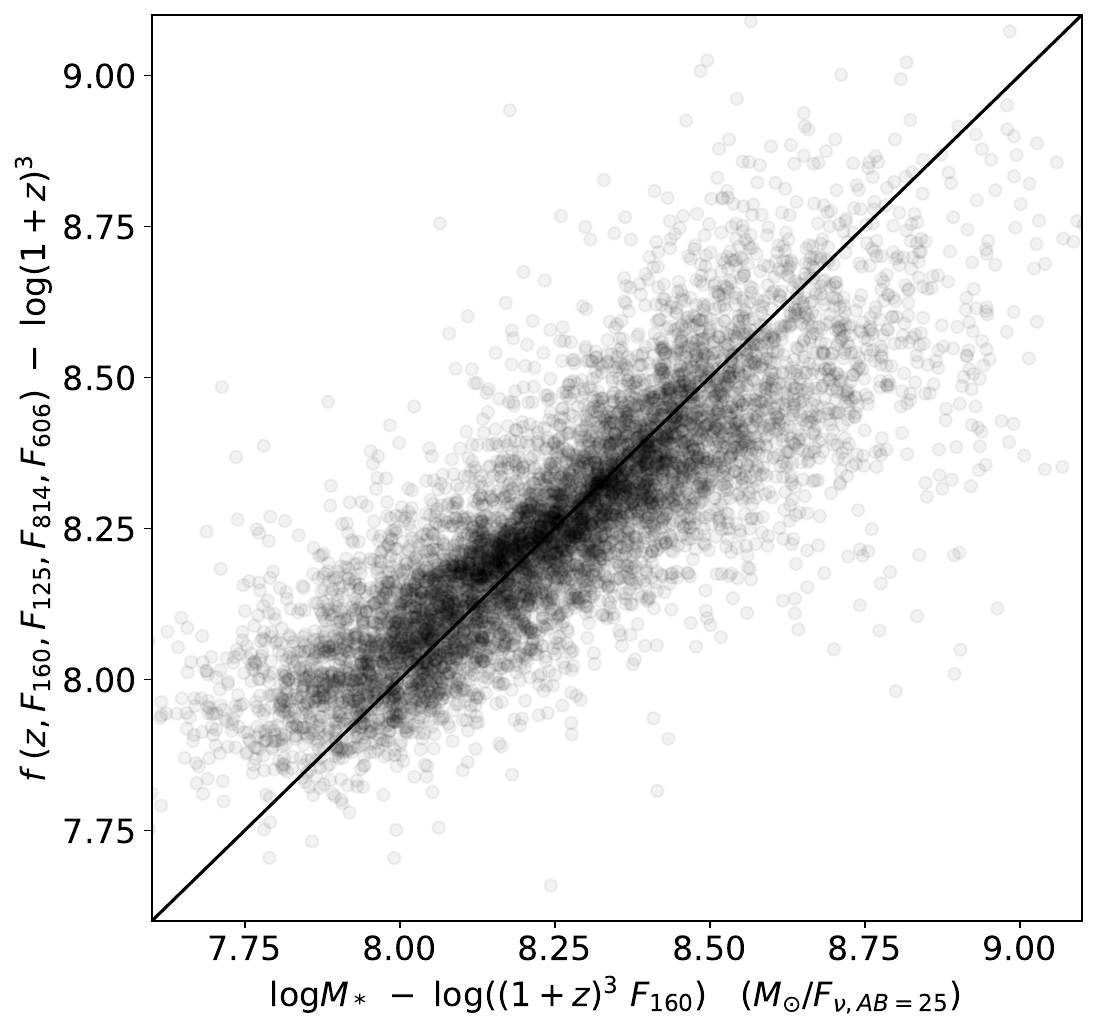}
\caption{ Correlation between the ground-truth mass-to-light ratio $\log{(M_\star/F_{160})}$ inferred from UV-to-infrared SED fitting) and our proxy $\log{(M_\star/F_{160})}$ based on four HST/ACS$+$WFC3-filter photometry as written in Eq.~1 and Table 1. A factor $(1+z)^3$, is removed as a trivial component in the cosmological distance dependence in the $F_{160}$ flux density.
\label{fig:color_ml}}
\end{figure}

We fit two separate relations for quiescent and star-forming galaxies with stellar masses $M_\star>10^{10}~M_{\odot}$ and over the redshift range $0.5<z<2.3$, beyond which only one data point redward of the Balmer/4000$\AA$~remains and the uncertainties in $M_\star/L$ estimates increase markedly.  We fit these relations to all galaxies with $M_\star>10^{10}~M_{\odot}$ and measured flux densities in the four HST filters and give the coefficients in Table 1. The resulting $\log{(M_\star/F_{160})}$~proxy is shown in Figure \ref{fig:color_ml}.  The overall scatter is 0.12 dex, increasing from 0.07 at $z<1$ to 0.15 at $z\sim 2$, which is less than or comparable to the typical uncertainty in SED-based $M_\star$~estimates. 

\begin{figure}[!t]
\epsscale{1.15}
\plotone{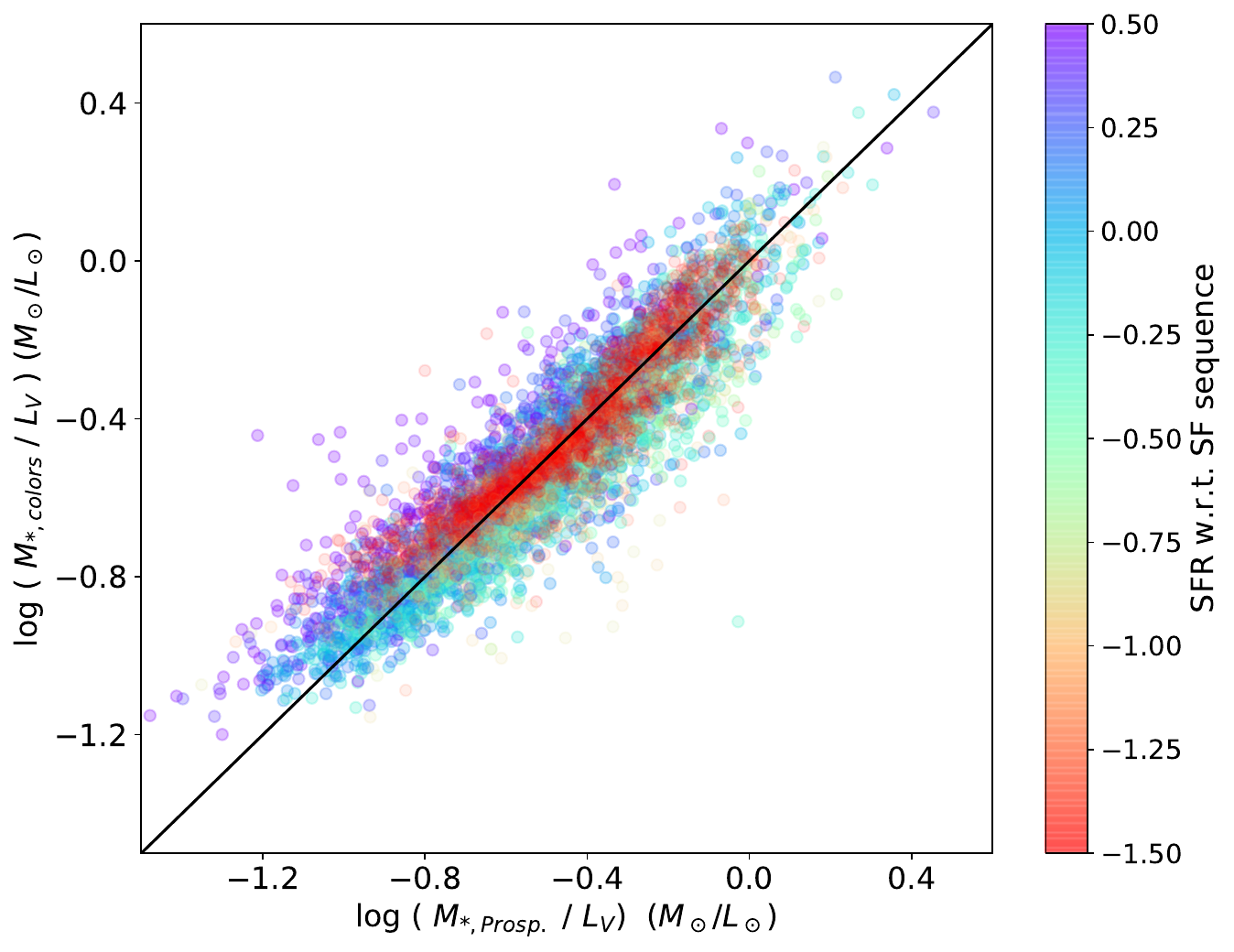}
\caption{ Comparison of rest-frame $V$-band stellar mass-to-light ratio as inferred from Prospector (x-axis) and our HST color-based proxy (y-axis). Across a 1.5 dex range in $M/L$ the HST color-based $M/L$ estimates agree well with the ground truth as inferred from fits to the full SED. The color-coding is the star-formation rate relative to the star-forming sequence defined by \citet{leja22}.
\label{fig:ml_ml}}
\end{figure}

\begin{figure*}[!t]
\epsscale{1.15}
\plotone{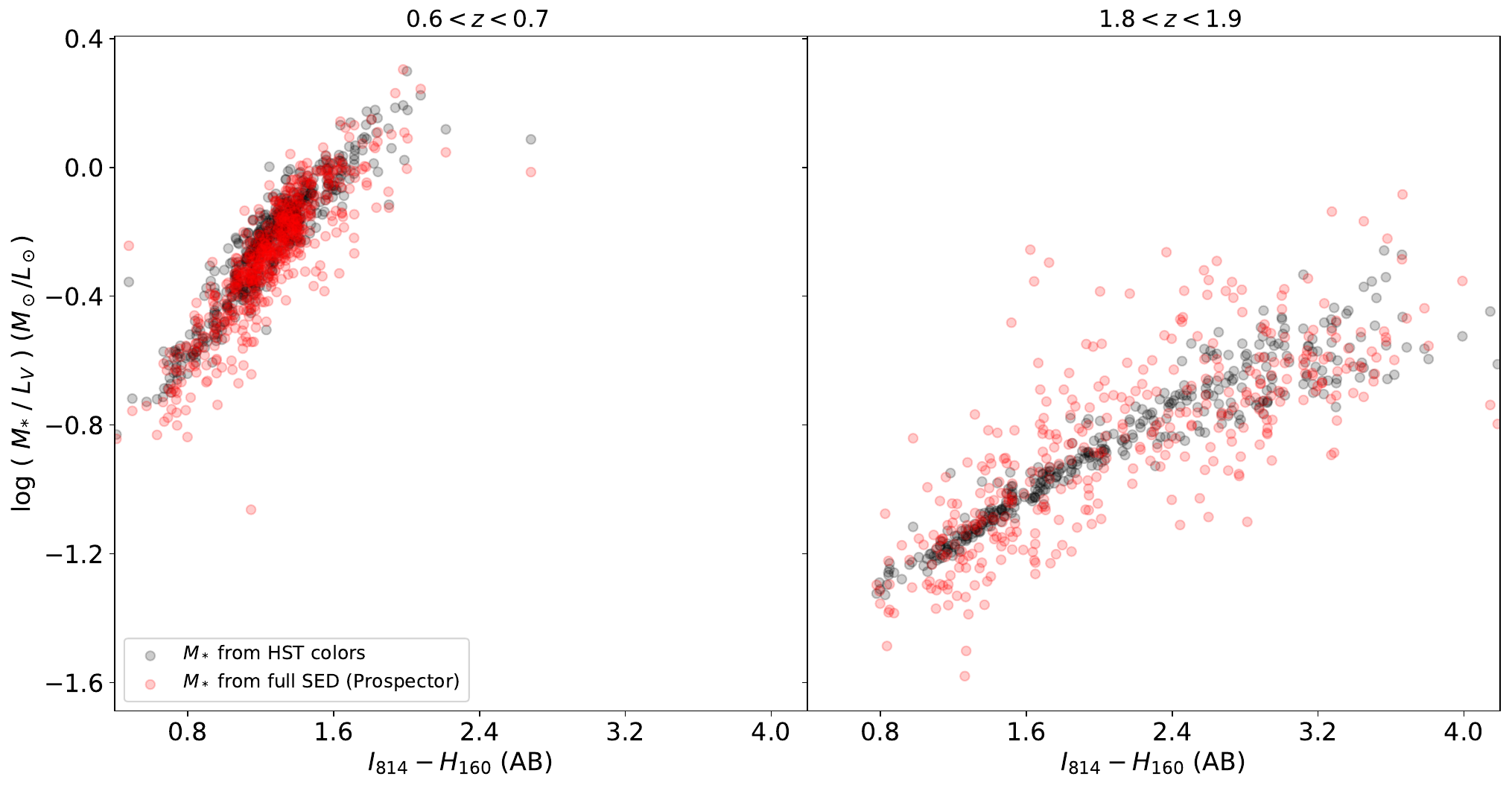}
\caption{ Color-$M/L$ relations for two narrow redshift bins, with in black the ground truth as inferred from full SED fits with Prospector and in red the HST color-based stellar $M/L$. The single relation in Eq.~1 produces different color-$M/L$ relations at different redshifts.
\label{fig:IH_ml}}
\end{figure*}

The functioning of the method is further illustrated in Figures \ref{fig:ml_ml} and \ref{fig:IH_ml}. Figure \ref{fig:ml_ml} shows a intuitively clearer version of Figure \ref{fig:color_ml}, displaying $M_\star/L$~values in the rest-frame $V$ band. The tightness of the correlation across a large dynamic range demonstrates the general precision and accuracy of our method. Underlying this result is an observed-frame color-$M/L$ relation that continuously changes with redshift (see Eq.~1). 

In Figure \ref{fig:IH_ml} we show for two redshift bins the relationship between $I_{\rm{814}}-H_{\rm{160}}$ and $M/L$. The $I_{814}-H_{160}$ is one of three colors used to derive the color-based $M_\star/L$ and the scatter is due to additional color information from $V_{606}$ and $J_{125}$. At low redshift the set of colors contains sufficient information to reproduce the variety in SED-based $M/L$ values at fixed color whereas at high redshift this is no longer the case due to a lack of information. 
If we fit a single relation to the joint population of star-forming and quiescent galaxies the distribution of points in Figure \ref{fig:ml_ml} becomes somewhat non-linear for quiescent galaxies, perhaps because those are greatly outnumbered by star-forming galaxies at higher redshifts.

Despite the apparent robustness of the method there are several caveats we have to keep in mind. The color-coding in Figure \ref{fig:ml_ml} reveals a remaining systematic effect: for star-forming galaxies there is a stratification with star-formation activity in the sense that the HST colors overpredict $M_\star/L_V$ for galaxies with high star-formation activity, and \textit{vice versa}. This does not translate into a stratification with attenuation; a degeneracy of stellar population properties must exist for a given (set of) colors. To what extent this issue affects the estimates of the stellar-half mass radius will be discussed where relevant.

Additional conceptual caveats are the following.
First, the Prospector stellar mass estimates serve as ground truth for our approach, but this ground truth itself is uncertain.  We test for the sensitivity to this particular choice by comparing with the original 3D-HST stellar mass estimates presented by \citet{skelton14}. Even though the fitted parameters and resulting color-$M/L$ relations change, the results after applying them to the observed color gradients as explained below do not differ significantly.  Second, the work by \citet{zibetti09} demonstrated that average $M/L$ estimates from integrated photometry can be biased due to the outshining effect of young, unobscured regions. This bias propagates into our color-$M/L$ relations and, more importantly, the impact on interpreting spatially resolved color information may differ.

\subsection{Converting Light Profiles to Stellar Mass Profiles}
\label{sec:m2d}

\subsubsection{Light Profile Fits and Color Gradients}
\label{sec:profiles}

\citet{nedkova21} describe the S\'ersic profile fits performed with {\tt galfitM} \citep{haussler13} on CANDELS imaging. The fits are performed on all available HST images in different filters simultaneously, fitting some parameters (axis ratio, position angle, position) as a constant while allowing others to vary quadratically as a function of wavelength (magnitude, effective radius, S\'ersic index). Uncertainties on the parameters are usually underestimated and we increase the uncertainties as prescribed by \citet{van-der-wel12}, who compared independent parameter estimates for the same objects and derived the `true' uncertainties as a function of $S/N$. 

The quadratic functions that describe the variation of the parameters with wavelength allow us to calculate S\'ersic flux density profiles as a function of radius $r$ at a common rest-frame wavelength of, e.g., 0.5$\mu$m ($S_{0.5}(r)$), with a half-light radius $R_{0.5}$ and a S\'ersic index $n_{0.5}$. The ratios of S\'ersic profiles at different wavelengths produce color profiles and can be used to define a color gradient between $0.5R_{0.5}$~and $2R_{0.5}$:

\begin{equation}
    \Delta C = \frac{S_{0.6}(2R_{0.5})~/~S_{0.4}(2R_{0.5})}{S_{0.6}(0.5R_{0.5})~/~S_{0.4}(0.5R_{0.5})}
    \label{eq2}
\end{equation}

The choices to evaluate the S\'ersic profiles at 0.4$\mu$m and 0.6$\mu$m, and between 0.5 and 2 effective radii, are motivated only by pragmatic considerations: all galaxies in the sample have this wavelength coverage and at smaller and larger radii the profiles are more uncertain. Through sampling the uncertainties in the profile fits we infer propagated uncertainties in the color profiles. 

\begin{figure*}[!t]
\epsscale{1.15}
\plotone{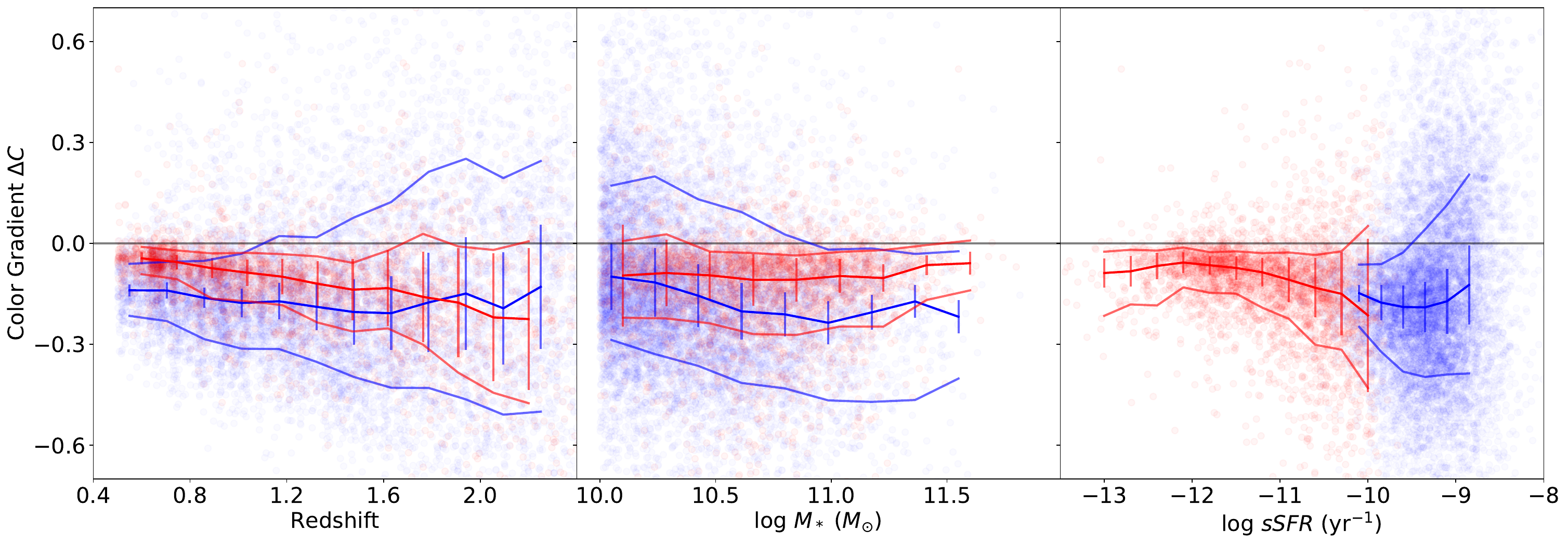}
\caption{Color gradient as defined in Eq.~2 vs.~redshift (left-hand panel), stellar mass (middle panel), and specific SFR (right-hand panel). The lines are running 16\%th, 50\%th, and 84\%th percentiles. Red and blue points and lines correspond with quiescent and star-forming galaxies, respectively. The error bars reflect the running median uncertainties for individual galaxies. The uncertainties are always smaller than the scatter, implying that we see intrinsic variations in the color gradient among galaxies. 
\label{fig:color_gradient}}
\end{figure*}

We note that the color gradient $\Delta C$ is not used in our method to derive $M_\star/L$~profiles and only serves to illustrate the strength of color gradient as a function of various galaxy parameters: Figure \ref{fig:color_gradient} shows the evolution of the color gradient and its dependence on stellar mass and star-formation activity. Negative color gradients (redder centers; bluer outerparts) are ubiquitous, at all $z$~and for all galaxy types. At $z\sim 1$ the measurement uncertainties are smaller than the population scatter, while at $z\sim 2$ they are similar, implying that at lower $z$~we can distinguish galaxies with different color gradients while at higher $z$ the observed scatter is dominated by measurement uncertainties. At fixed stellar mass star-forming galaxies generally have stronger gradients than quiescent galaxies.

\subsubsection{$M/L$ Gradients and Mass Profiles}
\label{sec:conversion}

The multi-color profiles are converted to $M/L$ profiles using the color-$M/L$ relations described in Section \ref{sec:colorml}. At each radius (within an elliptical annulus) we have a measured value of $F_{160}$, which is multiplied by the right-hand side of Eq.~1, with $z$ the redshift of the galaxies and where the flux ratios are given by the ratios of the four S\'ersic profiles in same annulus. The result is a direct conversion of the F160W light profile into an $M_\star$ profile. Since the inferred mass is based on a linear combination of different, inter-dependent S\'ersic profiles, it matches a S\'ersic profile itself (usually to within 0.1\%); we refit a S\'ersic profile to the mass profile out to 2$\times$~the effective radius in F160W, propagating -- with a Monte Carlo simulation -- the uncertainties on the individual light profile estimates and the color-$M/L$ relation.

\begin{figure*}[!t]
\epsscale{1.15}
\plotone{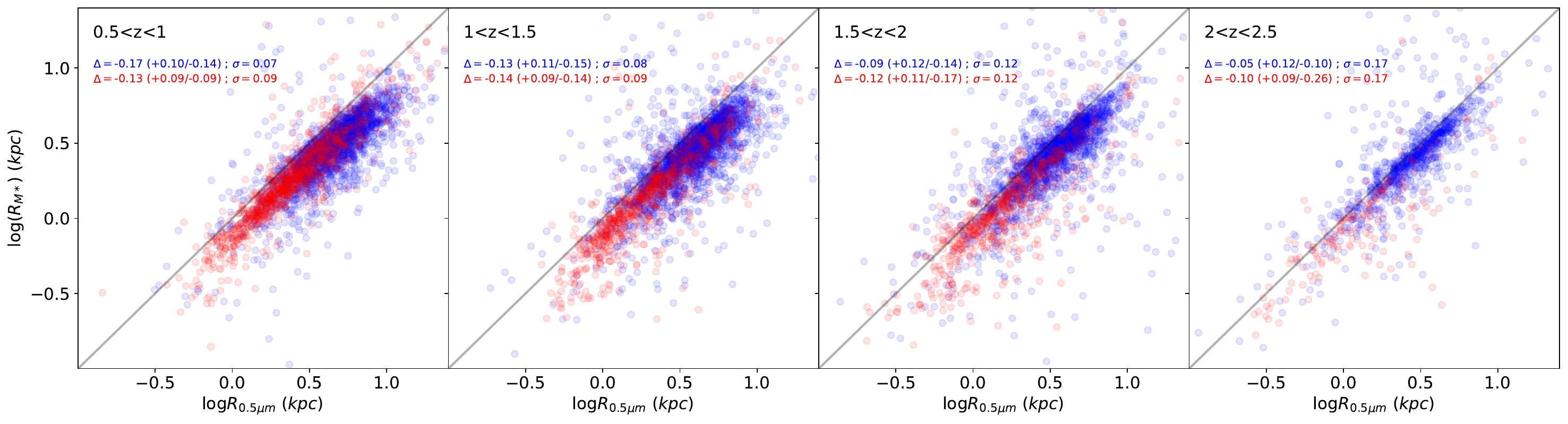}
\caption{ Stellar half-mass radius vs.~rest-frame optical half-light radius in four redshift bins for quiescent (red) and star-forming (blue) galaxies with total stellar mass $>10^{10}~M_\odot$. $\Delta$ is the median offset in dex, with the 16-84\%-ile scatter in parentheses. $\sigma$ is the median formal uncertainty. Offsets in the range $-0.05$~to $-0.17$ show that stellar half-mass radii are generally smaller than the optical radius. The scatter is comparable to the formal uncertainties, which implies that the uncertainties are certainly not underestimated.
\label{fig:rm_ropt}}
\end{figure*}

In Figure \ref{fig:rm_ropt} we show how the stellar half-mass radius \rmass~compares with the optical half-light radius \ropt. There is a generally tight correlation, with a scatter that increases from $\sim 0.10$~dex at $z<1$ to up to $\sim 0.2$~dex at $z>2$. Since the scatter is generally similar to the formal uncertainties, we conclude that uncertainties dominate over intrinsic variations at the level of the precision that we achieve. The exception is the set of star-forming galaxies at $z<1.5$, for which the scatter (0.1-0.15 dex) is somewhat larger than the uncertainties (0.08-0.09 dex); this implies that we recover intrinsic variations in  \rmass$/$\ropt from galaxy to galaxy (that are not explained by uncertainties). As expected based on the color gradients, \rmass~is generally smaller (by 0.1-0.15 dex) than \ropt, qualitatively consistent with previous work \citep{szomoru13, suess19a, suess19b, mosleh20, miller23}.  For quiescent galaxies we see an offset that is approximately constant with redshift ($-0.10$ to $-0.14$~dex across the entire redshift range), while for star-forming galaxies the difference decreases somewhat, from -0.17 dex at $z<1$ to -0.05 dex at $z>2$. At $z<1$~the largest galaxies are the most offset, which is due to the combination of a mild dependence on both $M_\star$~and $R$ (see Sec.~\ref{sec:sfq}).

It is quite remarkable that, overall the color and $M/L$ gradients are similar for star-forming and quiescent galaxies, as those in the former are mainly caused by a radial variation in attenuation \citep{miller22, matharu23}, even though, especially at lower redshifts, stellar population gradients are also present \citep{bell00}, while the latter generally have little dust and the gradient must be primarily due to stellar population variations.

The increase in scatter and downturn for galaxies with \ropt$\lesssim 1$~kpc suggests that systematic errors affect the \rmass~estimates for the smallest galaxies. This is not surprising, since the HST/WFC3 PSF has a FWHM of $\sim 1.4$~kpc. While light profiles can be well constrained at smaller scales, given sufficient S/N and accurate knowledge of the PSF, combining those from four different filters can lead to highly non-linear compound uncertainties that are difficult to propagate formally. Indeed, our formal uncertainties do not increase in line with the increased scatter. In Section \ref{sec:nircam} we will address the precision and accuracy of our \rmass~estimates, including the behavior at \ropt$\lesssim 1$~kpc.

\subsection{Comparison with NIRCam Effective Radii}
\label{sec:nircam}

\begin{figure*}[!t]
\epsscale{1.15}
\plotone{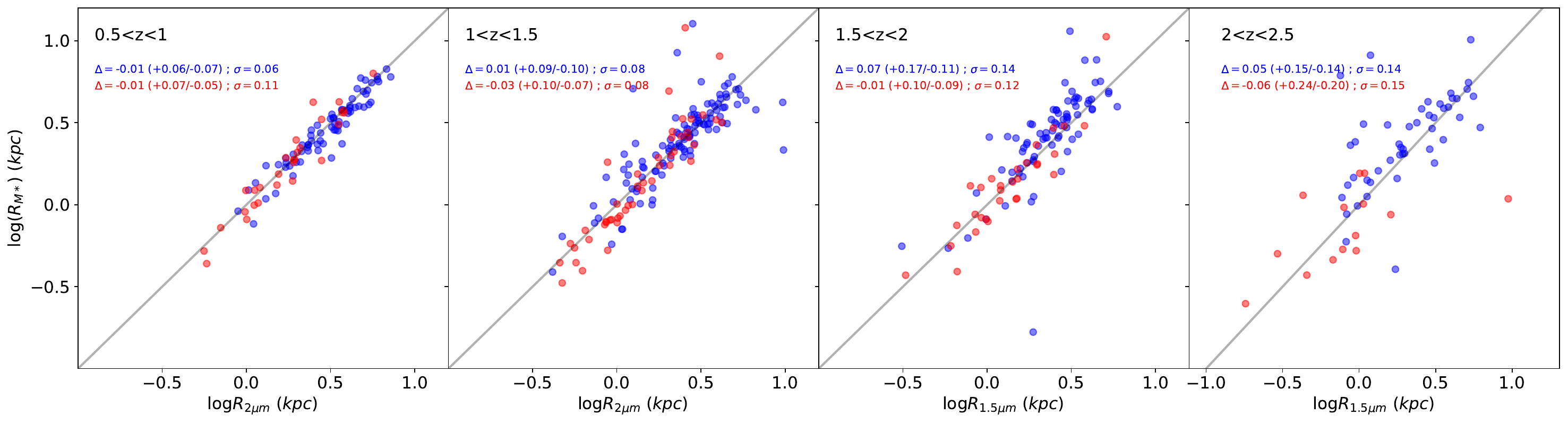}
\caption{ Stellar half-mass radius vs.~rest-frame near-infrared half-light radius derived from JWST/NIRCam imaging in four redshift bins for quiescent (red) and star-forming (blue) galaxies with total stellar mass $>10^{10}~M_\odot$. $\Delta$ is the median offset in dex, with in parentheses the 16-84\%-ile scatter. $\sigma$ is the median formal uncertainty. The lack of offsets and the similarity between scatter and formal uncertainty suggest that our \rmass~estimates do not suffer from biases and have reliable uncertainties.
\label{fig:rm_rnir}}
\end{figure*}

435 galaxies in our $M_\star>10^{10}~M_{\odot}$~sample ($\approx 4$\%) fall within the footprint of JWST/NIRCam imaging from CEERS \citep{finkelstein23}. Martorano et al.~(submittted) describe the {\tt galfitM} \citep{haussler13} fitting procedure to the 7 short- and long-wavelength filters images and the estimation of rest-frame near-IR S\'ersic profiles and associated half-light radii. After fitting the individual images a Chebychev polynomal fit is used to calculate the interpolated half-light radii at the desired rest-frame wavelength (0.5, 1.5, or 2.0$\mu$m). Relevant for the current paper is that we find no offset between the \ropt~estimates from CANDELS and CEERS ($<0.01$~dex) and small scatter ($\sim$0.05~dex).

In Figure \ref{fig:rm_rnir} we compare the stellar half-mass radii described in Section \ref{sec:profiles} with the NIRCam-based near-infrared half-light radii (rest-frame 2.0$\mu$m for $z<1.5$; rest-frame 1.5$\mu$m for $z>1.5$). We verified that this sub-sample is representative of the full CANDELS sample in terms of its \ropt~distribution: the statistical properties of the sub-sample are not significantly different from those shown in Fig.~\ref{fig:rm_ropt}.

Up to $z=1.5$ we see negligible offsets ($\leq 0.03$~dex), implying an absence of significant systematic biases in our stellar half-mass radii.  The typical formal uncertainty on the stellar half-mass estimates ($\leq 0.10$~dex) is very similar to the scatter, implying robust uncertainties and a typical level of precision of 25\% or better. For quiescent galaxies at $1.5<z<2$  the performance is still good, without a systematic offset, and uncertainties and scatter that are in agreement, while at $z>2$ the uncertainties increase and we are hampered by the small sample size as well. Reversing the question, we can also conclude that rest-frame near-IR light profiles represent stellar mass profiles well. That this is true is not immediately obvious, as age gradients would still cause $M/L$ gradients even in the near-IR. Either these effects are small or in the case of massive, high-metallicity galaxies this trend may be countered by a anti-correlation between metallicity and near-IR $M/L$.

For star-forming galaxies we see a small bias ($+0.05-0.07$~dex) that is comparable in magnitude but opposite in sign to the offset with half-light radius (Fig.~\ref{fig:rm_ropt}). This may imply that the stellar half-mass radii for star-forming galaxies are perhaps overestimated at $z>1.5$. Such issues are understandable, as it is challenging to obtain accurate $M/L$ estimates with limited photometric information redward of the 4000$\rm{\AA}$~break. The small bias in the color-$M_\star/L$ relation for star-forming galaxies identified in Section \ref{sec:colorml} (particularly, Fig.~\ref{fig:ml_ml}) may explain this: if galaxies have a positive gradient in sSFR \citep[e.g.,][]{tacchella17}, then their outer parts will have overesimed $M_\star/L$ with our method, leading to overestimated \rmass.
Clearly, NIRCam-based photometry can alleviate these concerns, but modeling of NIRCam photometry this is beyond the scope of this paper.

Unfortunately, the current CEERS NIRCam sample is too small to assess the robustness of \rmass~estimates at $<$1 kpc. In the $1<z<1.5$ panel there is a hint that those are indeed somewhat underestimated as suggested by Fig.~\ref{fig:rm_ropt}, if perhaps not by the same amount. The model NIRCam PSF is known to be inaccurate at some level and further progress in our understanding of the true NIRCam PSF is required for accurate size estimates of the smallest galaxies. We note that both the rest-frame optical sizes from HST and rest-frame near-IR sizes from JWST are based on imaging data with similar resolution ($\approx 0.15$~arcsec). Our statements regarding the precision and accuracy of our \rmass estimates are limited to the $>$1 kpc regime.

\subsection{Converting 2D Profiles to 3D Profiles}
\label{sec:deproject}
The methodology developed by \citet{van-de-ven21} allows us to convert our two-dimensional (projected) S\'ersic light and mass profiles into three-dimensional profiles. The procedure builds on our (statistical) knowledge of the intrinsic shape distribution of galaxies as described by \citet{chang13a, zhang19}: these authors constructed models for the projected shape distributions of galaxies of different types, masses and at different redshifts. These models assume a Gaussian distribution for the intrinsic axis ratios of a triaxial ellipsoid $c/a$ (short-to-long axis ratio) and $b/a$ (middle-to-long axis ratio) and/or a Gaussian distribution for the triaxiality parameter $T = (a^2-b^2)/(a^2-c^2)$. Depending on the type of galaxy the model consists of a single oblate population ($a\equiv b$), a single triaxial population, or a mixed model of two components (one oblate $+$ one triaxial).

Given these models, the redshift, stellar mass and star-formation activity of a galaxy produce an \textit{a priori} probability distribution for its intrinsic shape, parameterized as truncated Gaussian distributions for ellipticity ($E$) and triaxilaty ($T$) (as published in Table 3 of \citet{chang13a} and Table 1 of \citet{zhang19}), and its projected shape $q' = b'/a'$ (assuming random viewing angles for the intrinsic shape distribution). Then, given the measured projected shape $q'$ from the CANDELS imaging (see Sec.~\ref{sec:profiles}),
an \textit{a posteriori} probability distribution for the intrinsic shape distribution is constructed \citep[Eq.~9 from][]{van-de-ven21}: for a given $q$ exists a set of combinations of intrinsic shape and viewing angle. Instead of calculating this \textit{a posteriori} probability distribution for each galaxy we construct a library of solutions because its calculation requires an inversion of a demanding numerical integral and we wish to sample from the measurement uncertainty in $q$ and $R$ in order to propagate this into the inferred constraint on the intrinsic shape and 3D size.  Using these libraries we infer posterior probability distributions for the 3D profile and the associated parameters $R_{3D,M*}$, the radius of a sphere that contains 50\% of the triaxial stellar mass distribution.

\begin{figure}[!t]
\epsscale{1.15}
\plotone{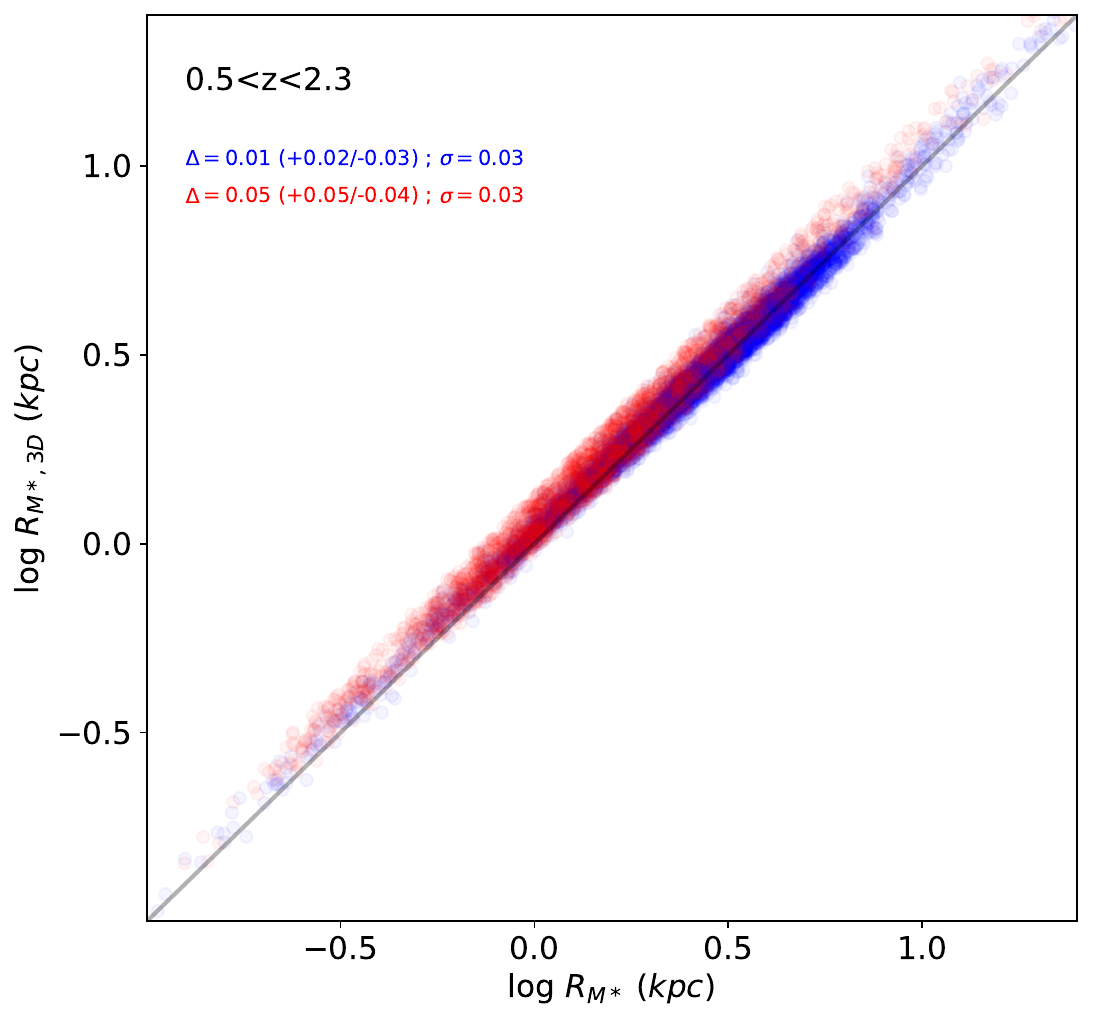}
\caption{
3D (deprojected) stellar half-mass radius vs.~2D (projected) stellar half-mass radius for quiescent (red) and star-forming (blue) galaxies with total stellar mass $>10^{10}~M_\odot$. $\Delta$ is the median offset in dex, with in parentheses the 16-84\%-ile scatter. $\sigma$ is the median formal uncertainty. 
\label{fig:rm_rm3d}}
\end{figure}

Figure \ref{fig:rm_rm3d} shows that the deprojection for galaxies in this mass range has a small effect on the inferred half-mass radii, as was already demonstrated for specific cases by \citet{van-de-ven21}. The sub-optimal visualisation of the results is chosen deliberately to highlight the lesser importance of deprojection compared to the $M_\star/L$~gradient correction shown in Figure \ref{fig:rm_ropt}. 

For quiescent galaxies, which show a larger variety in intrinsic shapes than star-forming galaxies (at least, for $M_\star>10^{10}~M_{\odot}$), the scatter in $R_{3D,M_\star}~ /~ R_{M_\star}$ is larger than for star-forming galaxies, but even for those the full range is no more than 30\%, with a systematic offset of $+$0.05 dex. For star-forming galaxies the scatter is smaller, and the systematic offset almost zero. In other words, the projected effective radius, measured as the long axis on the ellipse that encloses 50\% of the light or mass, generally serves as accurate and precise proxy for the median radius, the radius of a sphere that contains 50\% of the light or mass distribution in 3D. For low-mass star-forming galaxies, as shapes become more irregular, the deprojection will have a larger effect. Symmetrized uncertainties are defined as half of the 16th-84th percentile ranges of the posterior distributions, which produces a typical combined uncertainty in $R$ due to the deprojection of 0.03 dex ($\sigma$~in Fig.~\ref{fig:rm_rm3d}).

One caveat is that we have ignored the wavelength dependence on the shape. But given the lack of sensitivity to changes in intrinsic shape, we can be confident that this approximation does not affect the results. Another caveat is related to the finding by \citet{zhang19} that a correlation exists between (intrinsic) shape and size for star-forming galaxies. This implies that size should, in principle, be included in our construction of the \textit{a posteriori} probability distribution for the 3D profile. We test for the necessity of this additional step by varying the models, shifting the Gaussian means by 2$\sigma$ up and down. The differences are negligible (on the 1\%~level), which implies that the current setup -- where we ignore the covariance between size and shape -- is sufficient for our purposes.\footnote{The corollary implication is that differences in shape as a function of, e.g., redshift and mass are not relevant for the conversion from 2D to 3D profile in the first place, but this was not a foregone conclusion.} Finally, the stellar masses, star-formation rates and definition of quiescence used in this paper are not the same as those used by \citet{chang13a} and \citet{zhang19}, but given the minor effects of the deprojection these differences have no impact on the overall result.

\section{The Size-Mass Distribution and its Evolution}
\label{sec:mr}

\subsection{Comparing Size Proxies}

\begin{figure*}[!t]
\epsscale{1.15}
\plotone{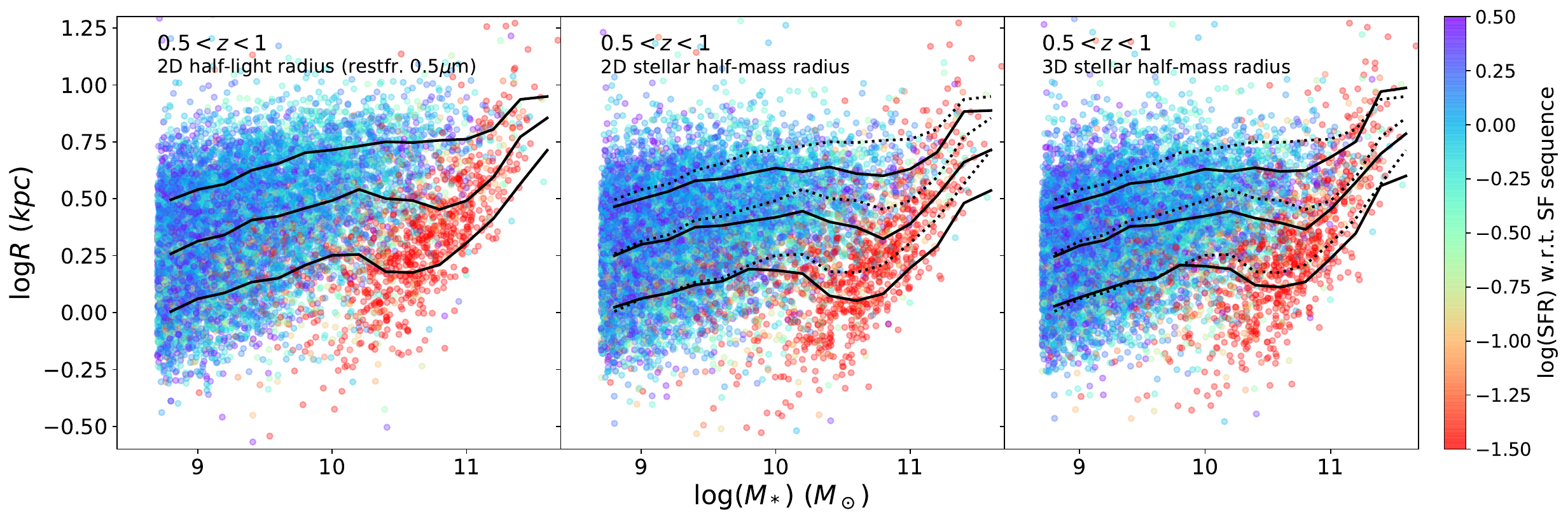}
\plotone{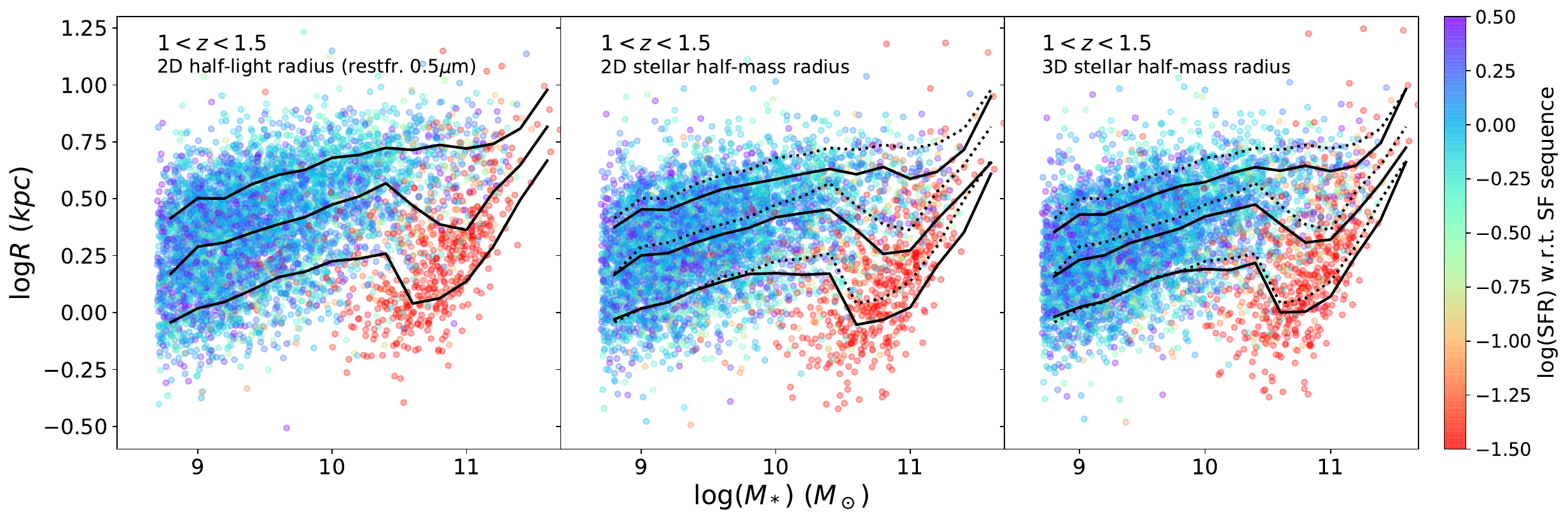}
\caption{ Size-stellar mass distributions for two redshift bins ($0.5<z<1$ at the top; $1<z<1.5$ at the bottom) and three different size proxies: rest-frame optical half-light radius $R_{\rm{opt}}$ (\textit{left}; projected stellar half-mass radius $R_{\rm{M_\star,2D}}$ (\textit{middle}); deprojected stellar half-mass radius $R_{\rm{M_\star,3D}}$ (\textit{bottom}). The solid lines indicate the 16th, 50th and 84th percentiles of the size distribution in 0.2 dex wide bins of $M_\star$ (Table \ref{tab:medians}). The dotted lines in the middle and right-hand panels repeat, for reference, the solid lines in the left-hand panels. The color-coding is the star-formation rate relative to the star-forming sequence defined by \citet{leja22}.
\label{fig:m_r}}
\end{figure*}

The significant but approximately constant offset between the rest-frame optical half-light radii and stellar half-mass radii, along with a lack of strong projection effects, imply that the view of the size-mass distribution of galaxies will not strongly depend on the choice of size proxy. In Figure \ref{fig:m_r} we show size-mass distributions for the redshift range $0.5<z<1.5$, for light-weighted radii and both projected (2D) and deprojected (3D) stellar half-mass radii. Regardless of size proxy we see the same characteristic size distribution, with a steep slope for quiescent galaxies and a shallower slope for star-forming galaxies. The combined median trend bends downward at $M_\star\approx 2\times 10^{10}~M_\odot$, a transitional point in the structural properties of present-day galaxies first identified by \citet{kauffmann03a}. The downward trend is particularly noticeable at $z>1$. 

Similarities aside, there are a number of small but interesting differences between the size proxies. Switching from light- to mass-weighted sizes strengthens the flattening/bending trend with mass further, an immediate result of stronger color and $M/L$ gradients seen for more massive galaxies. Projection effects play a relatively small role in the demographics of the size-mass distribution. But note that the dashed and solid median lines in the bottom-right panel of Fig.~\ref{fig:m_r} are nearly perfectly parallel. This implies that the joint effect of $M_\star/L$-gradient correction and deprojection leads to a constant shift in median galaxy sizes across more than 2 orders of magnitude in $M_\star$. The scatter in sizes decreases somewhat when performing these corrections: rather than increasing the overall error budget, the corrections get us closer to what can be considered true, physical sizes, here defined as 3D stellar half-mass radii. 

Increased uncertainties in the $R_{M_\star}$ estimates prevent us from presenting a similarly reliable view of the size-mass distribution at higher redshifts, in particular for star-forming galaxies. We should also keep in mind that the color-$M_\star/L$ relations devised in this paper are constructed on the basis of galaxies more massive than $M_\star > 10^{10}~M_\odot$,  therefore the \rmass~estimates of low-mass galaxies may be biased, but this issue is likely not important as a general absence of strong color gradients implies a general absence of strong $M_\star/L$ gradients. The other caveat is that we concluded in Section \ref{sec:profiles} that \rmass$\approx 1$ kpc size estimates are suspect if color gradients are present. The small-size tail of quiescent galaxies may therefore suffer from currently unknown systematic effects and increased random uncertainties. That said, the average sizes general exceed 2 kpc so that the trends shown in Figure \ref{fig:m_r} are robust.

\begin{deluxetable*}{ccccccccc}
\tabletypesize{\scriptsize}
\tablecolumns{12}
\tablecaption{Stellar Half-Mass Radii}
\tablehead{
\colhead{Field} &  \colhead{ID} &  \colhead{R.A.} &  \colhead{Dec.} & \colhead{z} & \colhead{$\log M_\star$} & \colhead{Quiescent Flag} & \colhead{$R_{M_\star}$} &  \colhead{$R_{M_\star,3D}$} \\
\colhead{} & \colhead{} & \colhead{deg.} & \colhead{deg.} & \colhead{}  & \colhead{$M_\odot$} & \colhead{} & \colhead{kpc} & \colhead{kpc}
}
\startdata
1  &  19     &  215.299759  &  53.051308   &  1.076  &  $9.510^{+0.076}_{-0.079}$   &  0  &  2.578$\pm$0.371    &  2.547$\pm$0.551     \\
1  &  28     &  215.264175  &  53.027222   &  0.763  &  $8.893^{+0.167}_{-0.142}$   &  0  &  1.840$\pm$0.234    &  1.746$\pm$0.324     \\
1  &  46     &  215.293457  &  53.048298   &  1.217  &  $10.650^{+0.031}_{-0.027}$  &  1  &  0.744$\pm$0.120    &  0.862$\pm$0.154     \\
1  &  55     &  215.298920  &  53.052399   &  0.697  &  $9.045^{+0.076}_{-0.088}$   &  0  &  2.557$\pm$0.294    &  2.564$\pm$0.496     \\
1  &  63     &  215.296555  &  53.050770   &  1.219  &  $9.931^{+0.049}_{-0.052}$   &  0  &  2.847$\pm$0.317    &  2.997$\pm$0.534     \\
1  &  83     &  215.302460  &  53.055332   &  0.725  &  $9.808^{+0.037}_{-0.043}$   &  0  &  2.422$\pm$0.449    &  2.665$\pm$0.633     \\
1  &  110    &  215.297195  &  53.051907   &  0.694  &  $10.294^{+0.049}_{-0.036}$  &  1  &  1.528$\pm$0.290    &  1.935$\pm$0.395     \\
1  &  145    &  215.281769  &  53.042686   &  0.784  &  $9.360^{+0.055}_{-0.160}$   &  0  &  4.058$\pm$0.504    &  4.034$\pm$0.796     \\
1  &  148    &  215.248138  &  53.019707   &  0.933  &  $8.947^{+0.086}_{-0.116}$   &  0  &  1.760$\pm$0.419    &  1.794$\pm$0.511     \\
1  &  151    &  215.252319  &  53.022339   &  0.679  &  $9.738^{+0.052}_{-0.062}$   &  0  &  2.300$\pm$0.224    &  2.159$\pm$0.332     \\
\vdots & \vdots & \vdots & \vdots & \vdots & \vdots & \vdots & \vdots & \vdots
\enddata
\tablecomments{(1): Field (1: EGS; 2: COSMOS; 3: GOODS-N; 4: GOODS-S; 5: UDS); (2): ID from \citet{skelton14}; (3): R.A.~from \citet{skelton14}; (4): Dec.~.from \citet{skelton14}; (5) Stellar mass from \citet{leja20}, with 16th- and 84th-perceintile uncertainty range; (6) Star-forming (0) or quiescent (1) (Sec.~\ref{sec:sed}); (7) Projected (2D) stellar half-mass radius (Sec.~\ref{sec:m2d}), defined as semi-major axis of half-mass ellipse; Deprojected (3D) stellar half-mass radius (Sec.~\ref{sec:deproject}), defined as radius of sphere containing 50\% of the stellar mass.} 
\label{tab:cat1}
\end{deluxetable*}

\subsection{Separating Star-Forming and Quiescent Galaxies}
\label{sec:sfq}

\begin{figure*}[!t]
\epsscale{1.15}
\plotone{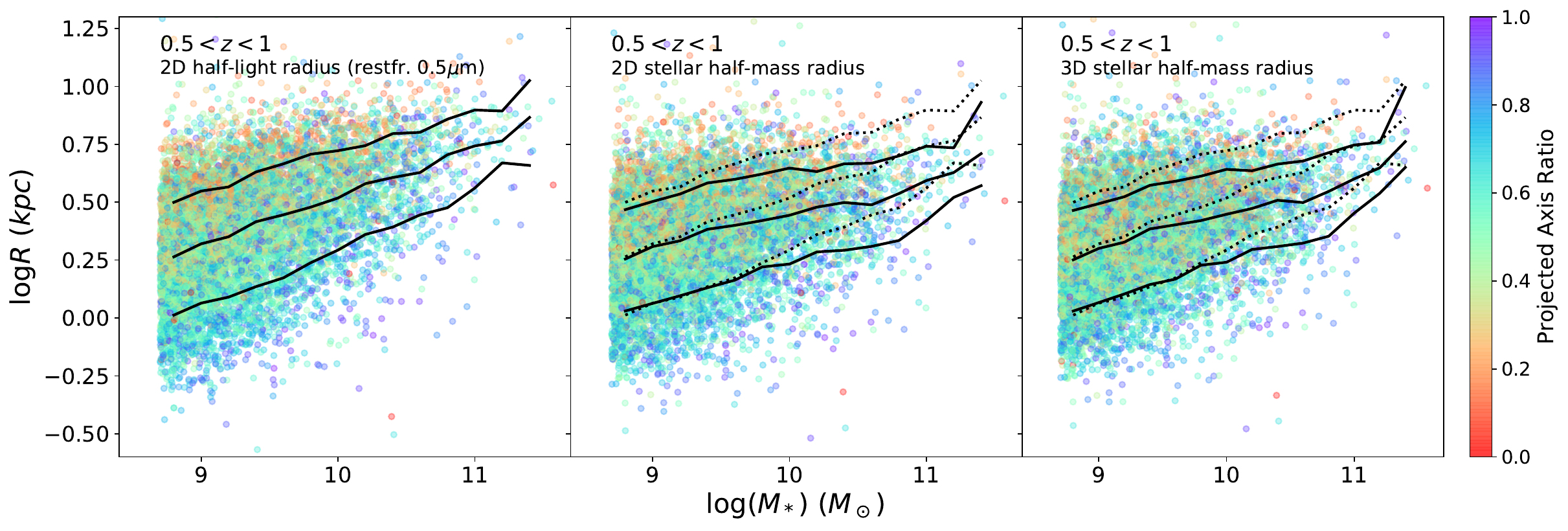}
\plotone{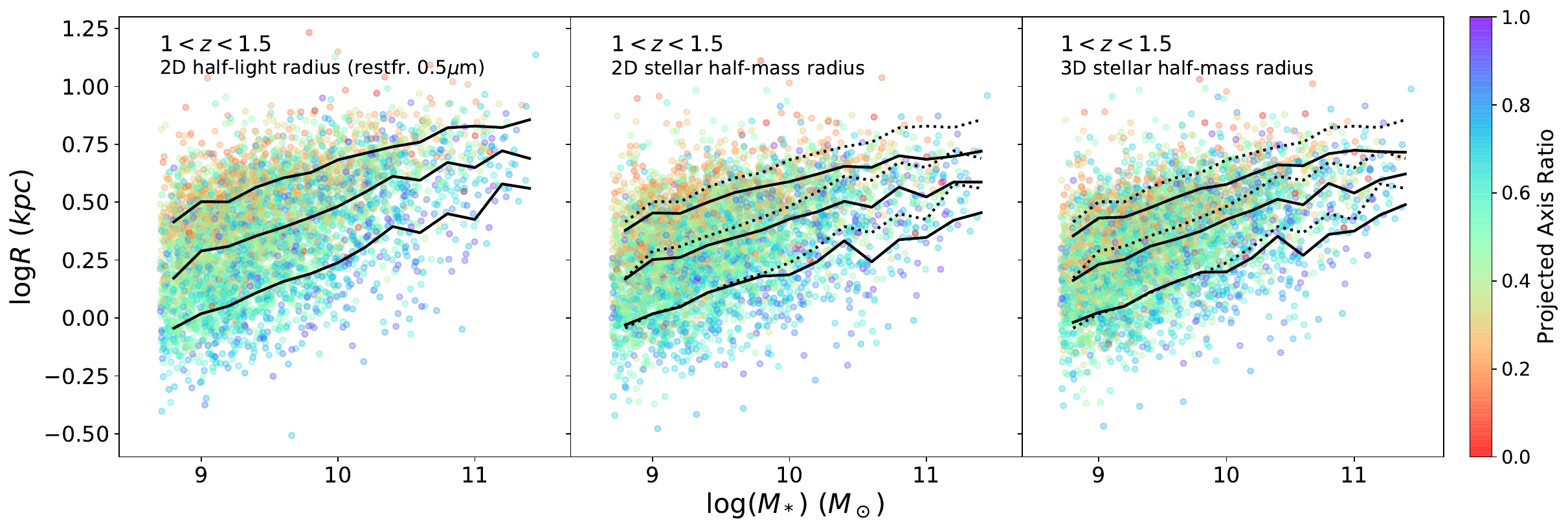}
\caption{ Size-stellar mass distributions of star-forming galaxies for two redshift bins ($0.5<z<1$ at the top; $1<z<1.5$ at the bottom) and three different size proxies: rest-frame optical half-light radius $R_{\rm{opt}}$ (\textit{left}; projected stellar half-mass radius $R_{\rm{M_\star,2D}}$ (\textit{middle}); deprojected stellar half-mass radius $R_{\rm{M_\star,3D}}$ (\textit{bottom}). The solid lines indicate the 16th, 50th and 84th percentiles of the size distribution in 0.2 dex wide bins of $M_\star$ (Table \ref{tab:medians_sf}). The dotted lines in the middle and right-hand panels repeat, for reference, the solid lines in the left-hand panels. The color-coding is the projected axis ratio.
\label{fig:m_r_sf}}
\end{figure*}

The clear pattern with star-formation activity in the size-mass distribution (Fig.~\ref{fig:m_r}) invites a closer look at the size distributions for star-forming and quiescent galaxies separately. Figure \ref{fig:m_r_sf} shows the star-forming galaxies and the most eye-catching result is the stellar-half mass radius depends less strongly on stellar mass than the half-light radius. Galaxies near the knee of the stellar mass function have \rmass~just $\approx 2$ times larger than galaxies 100$\times$ less massive, a slope of 0.15 dex. Also note the correlation between size and projected axis ratio, particularly at $M_\star<10^{10}~M_\odot$: as shown by \citet{zhang19}, galaxies that are flat in projection are more likely to have prolate/elongated 3D shapes and have larger (projected) sizes than galaxies with an oblate (disk-like) shape. When deprojecting the mass distribution this effect is lessened, but without a meaningful change in median size or scatter. This implies that for individual galaxies a shape-dependent deprojection correction can improve the accuracy of the size estimate, but that such a correction is not necessary to correctly infer the ensemble size distribution.

\begin{figure*}[!t]
\epsscale{1.15}
\plotone{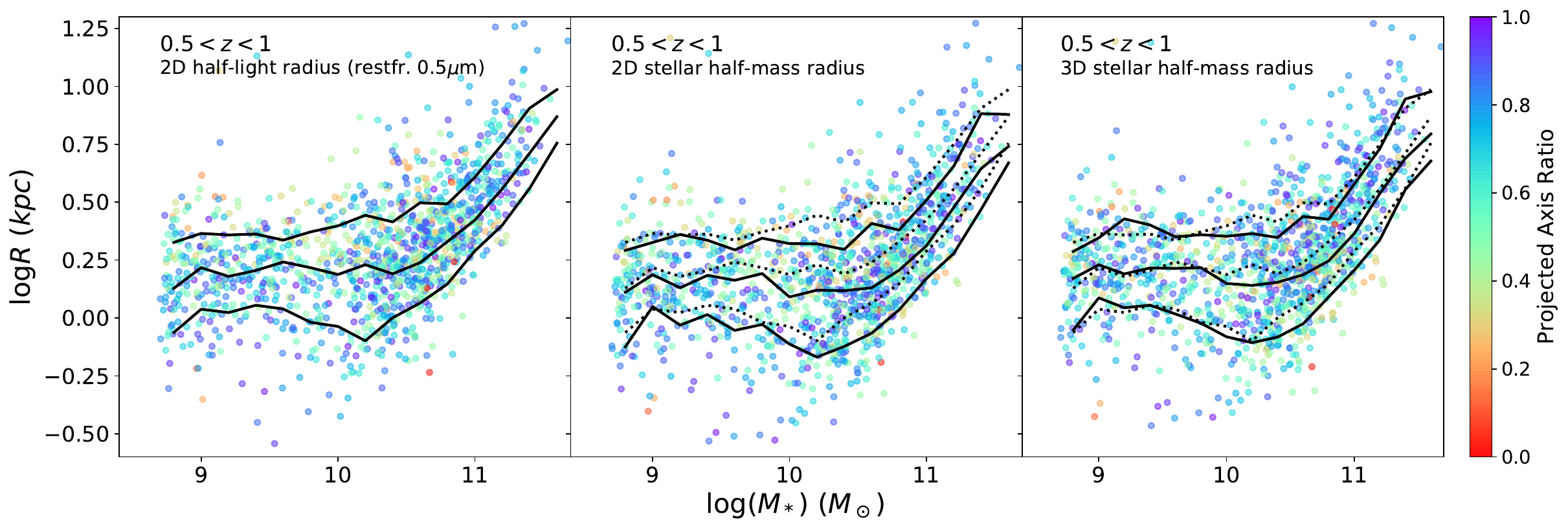}
\plotone{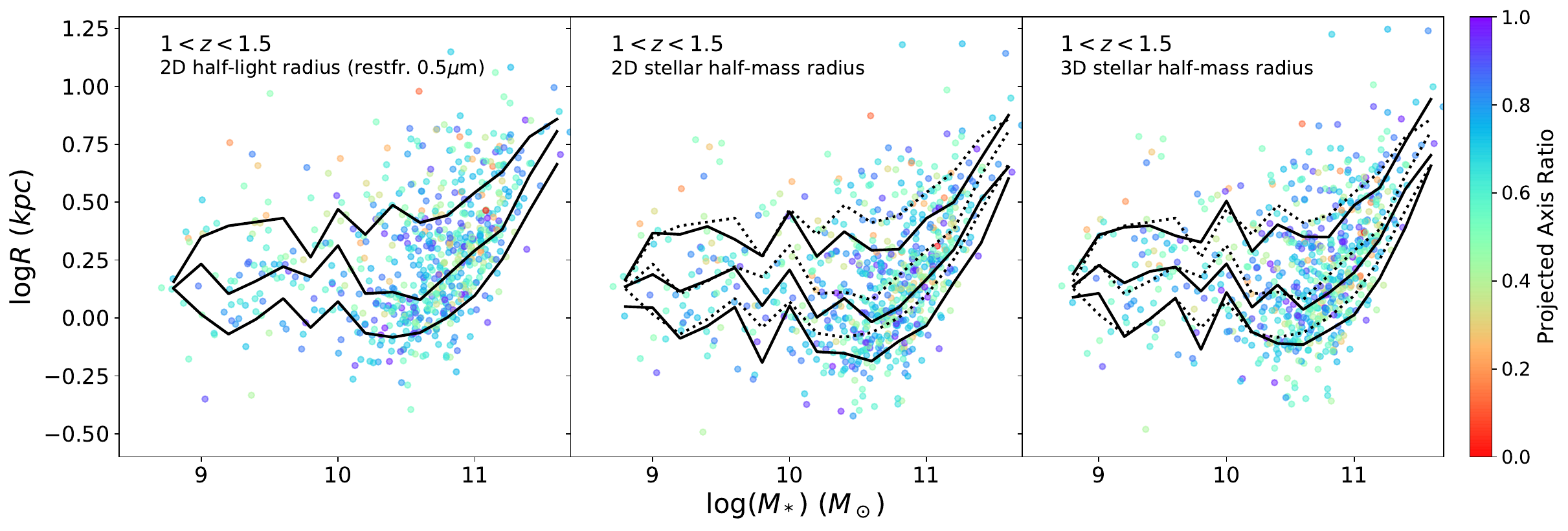}
\caption{ Size-stellar mass distributions of quiescent galaxies for two redshift bins ($0.5<z<1$ at the top; $1<z<1.5$ at the bottom) and three different size proxies: rest-frame optical half-light radius $R_{\rm{opt}}$ (\textit{left}; projected stellar half-mass radius $R_{\rm{M_\star,2D}}$ (\textit{middle}); deprojected stellar half-mass radius $R_{\rm{M_\star,3D}}$ (\textit{bottom}). The solid lines indicate the 16th, 50th and 84th percentiles of the size distribution in 0.2 dex wide bins of $M_\star$ (Table \ref{tab:medians_q}). The dotted lines in the middle and right-hand panels repeat, for reference, the solid lines in the left-hand panels. The color-coding is the projected axis ratio.
\label{fig:m_r_q}}
\end{figure*}

For quiescent galaxies (Fig.~\ref{fig:m_r_q}) the size-mass distribution is strikingly different from that of star-forming galaxies, but rather similar for the different size proxies. The $M_\star/L$~gradient correction shifts the sizes downward, partially countered by an upward shift when correcting for projection effects. The distribution is flat up until $\approx 2-3\times 10^{10}~M_\odot$ \citep[also see][]{nedkova21}, followed by a steep increase toward larger masses (seen by many authors). Correlations with projected axis ratio are less obvious compared to those seen for star-forming galaxies, as most galaxies are oblate or round/triaxial, but the smallest galaxies in the mass range $10^{10.5-11}~M_\odot$ are flatter (and therefore diskier) than average.

\begin{deluxetable*}{|c|ccc|ccc|ccc||ccc|ccc|ccc|}
\tabletypesize{\scriptsize}
\tablecaption{Median Radii and Percentiles\label{tab:medians}}
\tablehead{
	& \multicolumn{9}{c||}{$0.5<z<1$} & \multicolumn{9}{c|}{$1<z<1.5$} \\
     \cline{2-19}
	& \multicolumn{3}{c|}{$\log R_{0.5\mu\rm{m}}$} & \multicolumn{3}{c|}{$\log R_{M_\star}$} & \multicolumn{3}{c||}{$\log R_{M_\star,3D}$} & \multicolumn{3}{c|}{$\log R_{0.5\mu\rm{m}}$} & \multicolumn{3}{c|}{$\log R_{M_\star}$} & \multicolumn{3}{c|}{$\log R_{M_\star,3D}$}\\
        $\log M_\star~(M_\odot)$ & 16\% & 50\% & 84\%  & 16\% & 50\% & 84\%  & 16\% & 50\% & 84\%  & 16\% & 50\% & 84\%  & 16\% & 50\% & 84\%  & 16\% & 50\% & 84\% 
}
\startdata
8.8   &  0.01  &  0.26  &  0.50  &  0.03  &  0.25  &  0.47  &  0.03  &  0.25  &  0.46  &  -0.04  &  0.17  &  0.41  &  -0.03  &  0.17  &  0.37  &  -0.02  &  0.16  &  0.35  \\
9.0   &  0.06  &  0.31  &  0.54  &  0.06  &  0.30  &  0.50  &  0.07  &  0.29  &  0.49  &  0.02   &  0.29  &  0.50  &  0.02   &  0.25  &  0.45  &  0.02   &  0.23  &  0.43  \\
9.2   &  0.09  &  0.34  &  0.56  &  0.09  &  0.32  &  0.53  &  0.10  &  0.32  &  0.52  &  0.05   &  0.31  &  0.50  &  0.04   &  0.26  &  0.45  &  0.05   &  0.25  &  0.43  \\
9.4   &  0.13  &  0.41  &  0.62  &  0.12  &  0.37  &  0.58  &  0.14  &  0.38  &  0.57  &  0.10   &  0.35  &  0.56  &  0.10   &  0.31  &  0.50  &  0.10   &  0.31  &  0.48  \\
9.6   &  0.15  &  0.42  &  0.65  &  0.14  &  0.38  &  0.59  &  0.15  &  0.38  &  0.58  &  0.16   &  0.39  &  0.60  &  0.14   &  0.34  &  0.54  &  0.15   &  0.34  &  0.52  \\
9.8   &  0.21  &  0.46  &  0.70  &  0.19  &  0.40  &  0.61  &  0.21  &  0.41  &  0.60  &  0.18   &  0.42  &  0.63  &  0.17   &  0.37  &  0.56  &  0.18   &  0.37  &  0.56  \\
10.0  &  0.25  &  0.49  &  0.71  &  0.19  &  0.42  &  0.63  &  0.20  &  0.42  &  0.63  &  0.23   &  0.47  &  0.68  &  0.17   &  0.42  &  0.59  &  0.19   &  0.42  &  0.57  \\
10.2  &  0.26  &  0.54  &  0.73  &  0.17  &  0.45  &  0.62  &  0.19  &  0.45  &  0.62  &  0.24   &  0.51  &  0.69  &  0.17   &  0.44  &  0.61  &  0.19   &  0.45  &  0.61  \\
10.4  &  0.18  &  0.50  &  0.75  &  0.07  &  0.40  &  0.64  &  0.12  &  0.41  &  0.64  &  0.26   &  0.57  &  0.72  &  0.17   &  0.45  &  0.63  &  0.22   &  0.47  &  0.64  \\
10.6  &  0.18  &  0.49  &  0.75  &  0.05  &  0.37  &  0.61  &  0.11  &  0.39  &  0.62  &  0.04   &  0.47  &  0.71  &  -0.05  &  0.36  &  0.61  &  0.00   &  0.39  &  0.62  \\
10.8  &  0.21  &  0.45  &  0.76  &  0.08  &  0.32  &  0.60  &  0.13  &  0.37  &  0.62  &  0.06   &  0.39  &  0.74  &  -0.03  &  0.26  &  0.64  &  0.00   &  0.31  &  0.64  \\
11.0  &  0.31  &  0.49  &  0.76  &  0.20  &  0.39  &  0.63  &  0.24  &  0.45  &  0.68  &  0.14   &  0.36  &  0.72  &  0.02   &  0.27  &  0.59  &  0.07   &  0.32  &  0.62  \\
11.2  &  0.41  &  0.59  &  0.81  &  0.29  &  0.51  &  0.70  &  0.35  &  0.57  &  0.75  &  0.29   &  0.53  &  0.74  &  0.21   &  0.41  &  0.62  &  0.25   &  0.44  &  0.65  \\
11.4  &  0.57  &  0.77  &  0.94  &  0.48  &  0.66  &  0.88  &  0.56  &  0.70  &  0.97  &  0.50   &  0.65  &  0.81  &  0.35   &  0.53  &  0.72  &  0.41   &  0.57  &  0.74  \\
11.6  &  0.71  &  0.85  &  0.95  &  0.54  &  0.71  &  0.89  &  0.60  &  0.79  &  0.99  &  0.67   &  0.82  &  0.98  &  0.61   &  0.66  &  0.95  &  0.66   &  0.72  &  0.98  \\
\enddata
\tablecomments{Percentiles correspond with the lines in Figure \ref{fig:m_r}.} 
\end{deluxetable*}

\begin{deluxetable*}{|c|ccc|ccc|ccc||ccc|ccc|ccc|}
\tabletypesize{\scriptsize}
\tablecaption{Median Radii and Percentiles for Star-Forming Galaxies\label{tab:medians_sf}}
\tablehead{
	& \multicolumn{9}{c||}{$0.5<z<1$} & \multicolumn{9}{c|}{$1<z<1.5$} \\
     \cline{2-19}
	& \multicolumn{3}{c|}{$\log R_{0.5\mu\rm{m}}$} & \multicolumn{3}{c|}{$\log R_{M_\star}$} & \multicolumn{3}{c||}{$\log R_{M_\star,3D}$} & \multicolumn{3}{c|}{$\log R_{0.5\mu\rm{m}}$} & \multicolumn{3}{c|}{$\log R_{M_\star}$} & \multicolumn{3}{c|}{$\log R_{M_\star,3D}$}\\
        $\log M_\star~(M_\odot)$ & 16\% & 50\% & 84\%  & 16\% & 50\% & 84\%  & 16\% & 50\% & 84\%  & 16\% & 50\% & 84\%  & 16\% & 50\% & 84\%  & 16\% & 50\% & 84\% 
}
\startdata
8.8   &  0.01  &  0.26  &  0.50  &  0.03  &  0.25  &  0.47  &  0.03  &  0.25  &  0.46  &  -0.04  &  0.17  &  0.41  &  -0.03  &  0.17  &  0.38  &  -0.02  &  0.16  &  0.35  \\
9.0   &  0.06  &  0.32  &  0.55  &  0.06  &  0.31  &  0.50  &  0.07  &  0.30  &  0.49  &  0.02   &  0.29  &  0.50  &  0.02   &  0.25  &  0.45  &  0.02   &  0.23  &  0.43  \\
9.2   &  0.09  &  0.35  &  0.57  &  0.10  &  0.33  &  0.54  &  0.10  &  0.33  &  0.52  &  0.05   &  0.31  &  0.50  &  0.05   &  0.26  &  0.45  &  0.05   &  0.25  &  0.44  \\
9.4   &  0.14  &  0.42  &  0.63  &  0.13  &  0.38  &  0.58  &  0.14  &  0.38  &  0.57  &  0.11   &  0.35  &  0.56  &  0.11   &  0.31  &  0.50  &  0.11   &  0.31  &  0.48  \\
9.6   &  0.17  &  0.45  &  0.67  &  0.16  &  0.40  &  0.60  &  0.17  &  0.40  &  0.59  &  0.16   &  0.39  &  0.60  &  0.15   &  0.35  &  0.54  &  0.16   &  0.34  &  0.52  \\
9.8   &  0.24  &  0.48  &  0.71  &  0.22  &  0.42  &  0.62  &  0.23  &  0.42  &  0.61  &  0.19   &  0.43  &  0.63  &  0.18   &  0.38  &  0.57  &  0.20   &  0.37  &  0.56  \\
10.0  &  0.29  &  0.52  &  0.72  &  0.23  &  0.44  &  0.65  &  0.24  &  0.45  &  0.64  &  0.24   &  0.48  &  0.68  &  0.19   &  0.43  &  0.59  &  0.20   &  0.43  &  0.58  \\
10.2  &  0.36  &  0.58  &  0.74  &  0.29  &  0.48  &  0.63  &  0.30  &  0.47  &  0.64  &  0.31   &  0.54  &  0.71  &  0.24   &  0.46  &  0.62  &  0.26   &  0.46  &  0.62  \\
10.4  &  0.39  &  0.61  &  0.80  &  0.29  &  0.50  &  0.67  &  0.31  &  0.51  &  0.66  &  0.39   &  0.61  &  0.74  &  0.33   &  0.50  &  0.65  &  0.35   &  0.51  &  0.66  \\
10.6  &  0.45  &  0.63  &  0.80  &  0.31  &  0.49  &  0.67  &  0.32  &  0.50  &  0.68  &  0.37   &  0.59  &  0.76  &  0.24   &  0.48  &  0.65  &  0.27   &  0.49  &  0.66  \\
10.8  &  0.48  &  0.70  &  0.86  &  0.33  &  0.54  &  0.70  &  0.35  &  0.55  &  0.71  &  0.45   &  0.67  &  0.82  &  0.34   &  0.56  &  0.70  &  0.36   &  0.58  &  0.71  \\
11.0  &  0.56  &  0.74  &  0.90  &  0.42  &  0.59  &  0.74  &  0.45  &  0.60  &  0.75  &  0.43   &  0.65  &  0.83  &  0.35   &  0.52  &  0.69  &  0.38   &  0.54  &  0.72  \\
11.2  &  0.67  &  0.76  &  0.89  &  0.52  &  0.63  &  0.73  &  0.54  &  0.65  &  0.76  &  0.58   &  0.72  &  0.82  &  0.42   &  0.59  &  0.70  &  0.44   &  0.60  &  0.72  \\
11.4  &  0.66  &  0.87  &  1.02  &  0.57  &  0.71  &  0.93  &  0.65  &  0.76  &  1.00  &  0.56   &  0.69  &  0.86  &  0.45   &  0.59  &  0.72  &  0.49   &  0.62  &  0.72  \\
\enddata
\tablecomments{Percentiles correspond with the lines in Figure \ref{fig:m_r_sf}.} 
\end{deluxetable*}

\begin{deluxetable*}{|c|ccc|ccc|ccc||ccc|ccc|ccc|}
\tabletypesize{\scriptsize}
\tablecaption{Median Radii and Percentiles for Quiescent Galaxies\label{tab:medians_q}}
\tablehead{
	& \multicolumn{9}{c||}{$0.5<z<1$} & \multicolumn{9}{c|}{$1<z<1.5$} \\
     \cline{2-19}
	& \multicolumn{3}{c|}{$\log R_{0.5\mu\rm{m}}$} & \multicolumn{3}{c|}{$\log R_{M_\star}$} & \multicolumn{3}{c||}{$\log R_{M_\star,3D}$} & \multicolumn{3}{c|}{$\log R_{0.5\mu\rm{m}}$} & \multicolumn{3}{c|}{$\log R_{M_\star}$} & \multicolumn{3}{c|}{$\log R_{M_\star,3D}$}\\
        $\log M_\star~(M_\odot)$ & 16\% & 50\% & 84\%  & 16\% & 50\% & 84\%  & 16\% & 50\% & 84\%  & 16\% & 50\% & 84\%  & 16\% & 50\% & 84\%  & 16\% & 50\% & 84\% 
}
\startdata
8.8   &  -0.06  &  0.13  &  0.33  &  -0.12  &  0.11  &  0.29  &  -0.05  &  0.17  &  0.29  &  0.13   &  0.13  &  0.16  &  0.05   &  0.14   &  0.17  &  0.09   &  0.14  &  0.19  \\
9.0   &  0.04   &  0.22  &  0.36  &  0.05   &  0.19  &  0.33  &  0.09   &  0.23  &  0.35  &  0.02   &  0.23  &  0.35  &  0.04   &  0.19   &  0.37  &  0.11   &  0.23  &  0.36  \\
9.2   &  0.02   &  0.18  &  0.36  &  -0.03  &  0.13  &  0.36  &  0.04   &  0.19  &  0.43  &  -0.07  &  0.10  &  0.40  &  -0.09  &  0.12   &  0.36  &  -0.08  &  0.15  &  0.39  \\
9.4   &  0.05   &  0.20  &  0.36  &  0.01   &  0.18  &  0.34  &  0.06   &  0.22  &  0.40  &  -0.01  &  0.16  &  0.41  &  -0.03  &  0.16   &  0.40  &  -0.00  &  0.20  &  0.40  \\
9.6   &  0.04   &  0.24  &  0.34  &  -0.05  &  0.16  &  0.29  &  0.02   &  0.21  &  0.35  &  0.08   &  0.22  &  0.43  &  0.05   &  0.22   &  0.34  &  0.09   &  0.22  &  0.36  \\
9.8   &  -0.02  &  0.22  &  0.37  &  -0.03  &  0.19  &  0.34  &  -0.02  &  0.22  &  0.36  &  -0.04  &  0.18  &  0.26  &  -0.19  &  0.05   &  0.27  &  -0.14  &  0.12  &  0.33  \\
10.0  &  -0.04  &  0.19  &  0.40  &  -0.11  &  0.09  &  0.32  &  -0.08  &  0.15  &  0.35  &  0.07   &  0.31  &  0.47  &  0.05   &  0.21   &  0.46  &  0.11   &  0.23  &  0.50  \\
10.2  &  -0.10  &  0.23  &  0.44  &  -0.17  &  0.12  &  0.32  &  -0.11  &  0.14  &  0.36  &  -0.07  &  0.11  &  0.36  &  -0.15  &  0.00   &  0.27  &  -0.06  &  0.05  &  0.29  \\
10.4  &  0.00   &  0.19  &  0.41  &  -0.12  &  0.12  &  0.30  &  -0.08  &  0.15  &  0.35  &  -0.08  &  0.11  &  0.49  &  -0.15  &  0.08   &  0.37  &  -0.11  &  0.14  &  0.40  \\
10.6  &  0.06   &  0.24  &  0.50  &  -0.07  &  0.13  &  0.41  &  -0.03  &  0.19  &  0.44  &  -0.06  &  0.08  &  0.41  &  -0.19  &  -0.02  &  0.29  &  -0.12  &  0.04  &  0.35  \\
10.8  &  0.15   &  0.33  &  0.49  &  0.03   &  0.21  &  0.38  &  0.09   &  0.25  &  0.43  &  0.00   &  0.18  &  0.44  &  -0.10  &  0.05   &  0.30  &  -0.05  &  0.11  &  0.35  \\
11.0  &  0.29   &  0.42  &  0.61  &  0.17   &  0.31  &  0.51  &  0.21   &  0.36  &  0.58  &  0.10   &  0.29  &  0.54  &  -0.03  &  0.17   &  0.43  &  0.01   &  0.20  &  0.49  \\
11.2  &  0.40   &  0.56  &  0.75  &  0.28   &  0.48  &  0.66  &  0.34   &  0.54  &  0.73  &  0.25   &  0.38  &  0.63  &  0.15   &  0.29   &  0.50  &  0.17   &  0.34  &  0.56  \\
11.4  &  0.56   &  0.71  &  0.91  &  0.47   &  0.64  &  0.88  &  0.55   &  0.69  &  0.95  &  0.46   &  0.61  &  0.78  &  0.32   &  0.51   &  0.69  &  0.39   &  0.56  &  0.76  \\
11.6  &  0.75   &  0.87  &  0.99  &  0.67   &  0.74  &  0.88  &  0.68   &  0.79  &  0.98  &  0.66   &  0.81  &  0.86  &  0.60   &  0.65   &  0.88  &  0.66   &  0.70  &  0.94  \\
\enddata
\tablecomments{Percentiles correspond with the lines in Figure \ref{fig:m_r_q}.} 
\end{deluxetable*}

\subsection{Size Evolution of Massive Quiescent Galaxies}
\label{sec:evol}

\begin{figure}[!t]
\epsscale{1.15}
\plotone{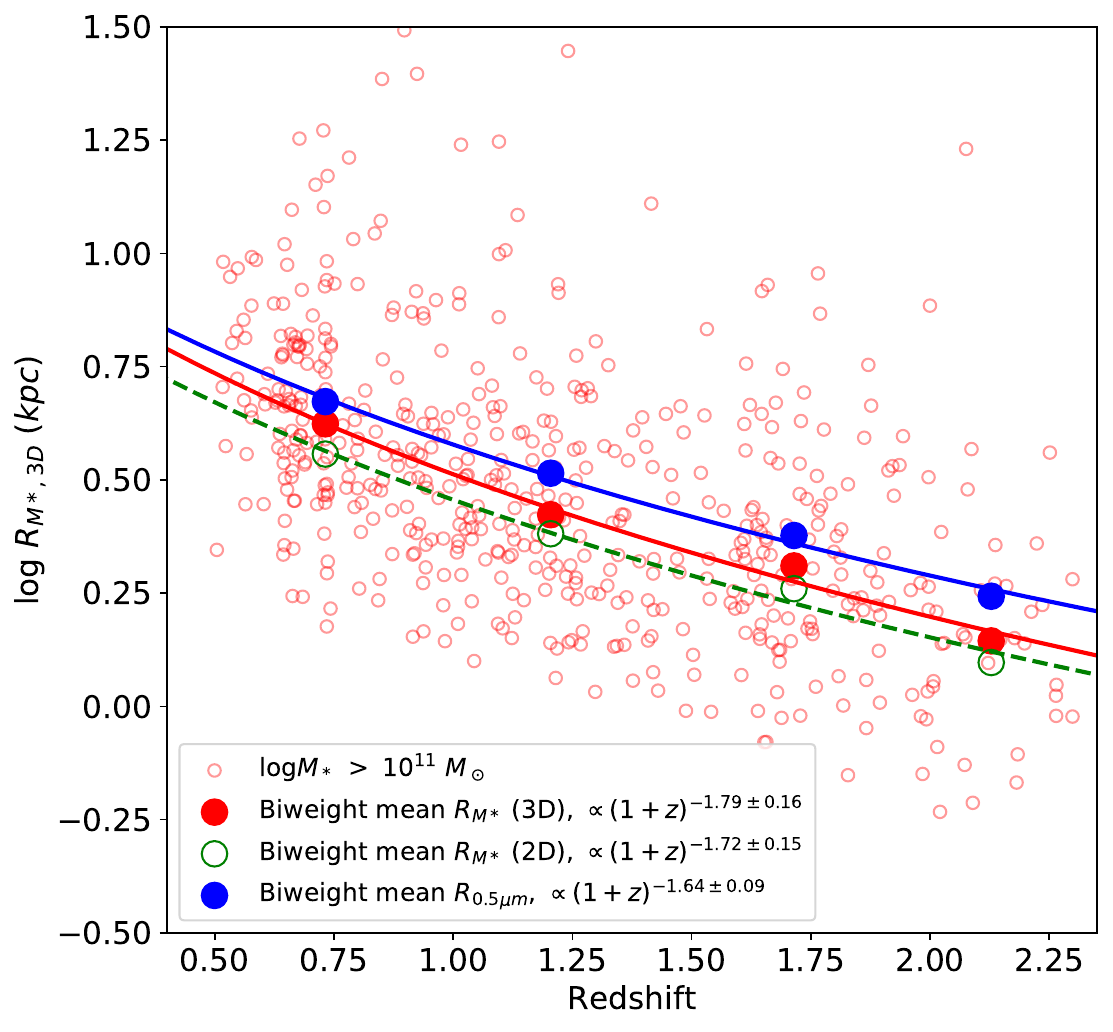}
\caption{ Stellar half-mass/light radius vs.~redshift for massive ($M_\star>10^{11}~M_\odot$) quiescent galaxies. The small data points are the 3D stellar half-mass radii for the individual objects, the large red symbols their mean values in bins of redshift. The open green symbols are the biweight mean values of the 2D stellar half-mass radii; the blue symbols the 2D stellar half-light radii at rest-frame 0.5$\mu$m. The uncertainties on the mean values are of the same order as the sizes of the symbols. The lines show that the pace of evolution, determined by simple regression fits, is always similar and in agreement with previously published results on the evolution of light-weighted sizes.
\label{fig:rz_massive}}
\end{figure}

For massive ($M_\star > 10^{11}~M_\odot$), quiescent galaxies our size estimates are robust at $z>1.5$ and we can probe the evolution of size-mass distribution out to $z=2.3$.\footnote{The data for this subset of galaxies is included in Table \ref{tab:cat1}.} Figure \ref{fig:rz_massive} shows the evolution with redshift of the sizes of massive, quiescent galaxies, comparing the three definitions: light-weighted ($R_{0.5\mu\rm{m}} \propto (1+z)^{-1.64\pm 0.09}$), mass-weighted ($R_{M_\star} \propto (1+z)^{-1.72\pm 0.15}$), and deprojected mass-weighted ($R_{M_\star,3D} \propto (1+z)^{-1.79\pm 0.16}$).  The key result is that the size evolution is significant, regardless of the choice of size proxy. At all redshifts the mass-weighted sizes are smaller than the light-weighted sizes by similar amounts by 0.1$-$0.15 dex, in line with the early NIRCam-based results from \citet{suess22} that showed that NIRCam/F444W sizes are smaller than NIRCam/F150W by $\approx 0.15$~dex. The deprojection from 2D to 3D (Sec.~\ref{sec:deproject}) shifts the sizes upward by about 0.05 dex. For most purposes, the measured sizes of massive galaxies do not require a deprojection correction to enable a meaningful comparison with the sizes of simulated galaxies based on 3D stellar particle distributions.

The above analysis does not consider the steep slope of the size-mass relation for massive galaxies, and the effect that evolution in slope or differences in slope among the three size proxies might have on the inferred evolution. But the similarity in slope for the three size proxies seen in Figure \ref{fig:m_r_q} and a lack of strong change in slope with redshift suggest that the effects are mild at most. Indeed, the size evolution result shown in Figure \ref{fig:rz_massive} does not depend on the precise selection in $M_\star$: galaxies with $10^{11}<M_\star/M_\odot < 2\times 10^{11}$ and galaxies with $M_\star>2\times 10^{11}~M_\odot$ show the same pace of evolution for all three size proxies, well within the uncertainties.

There are two effects that may introduce a bias the inferred size evolution. First, differences in slope among the size proxies  depend The slopes are very similar for the three size proxies (see Fig.\ref{fig:m_r_q}), so that any dependence on slope in the parameterization of size evolution is the same for all three proxies. A shift in the $M_\star$ distribution with redshift would introduce a bias in the estimated pace of evolution. Repeating the analysis by adopting a fixed slope of $\Delta \log{R}/\Delta \log{M_\star=0.6}$ does not change the pace of evolution by less than 25\% of the uncertainty.

Our result that the half-mass and half-light radii of massive quiescent galaxies evolve rapidly wiht redshift and in a similar manner (with $R\propto(1+z)^{\alpha=-1.6\dots -1.8}$) is in tension with previous work. \citet{suess19b, miller23} argue that a correction for $M_\star/L$ gradients removes much of the size evolution at $z>1$ seen in the rest-frame optical so that the stellar half-mass radius, on average, evolves much less than the stellar half-light radius or not at all.

This begs the question how the previously published stellar half-mass radii compare with the rest-frame near-IR sizes from NIRCam. In Appendix A we provide and extensive and quantitative comparison, the result of which is, in short, that the \rmass~estimates constructed in this paper produce the smallest offset and scatter when compared to NIRCam-based sizes.

In addition, for our half-mass radii the uncertainties are similar to the scatter in the comparison with the NIRCam sizes (typically, 0.10 dex at $z<2$; also see Fig.~\ref{fig:rm_rnir}), whereas for previously published estimates the formal uncertainties are smaller ($\lesssim 0.05$~dex) and likely underestimated, as was already pointed out by \citet{miller23}. The accurate agreement over the redshift range $0.5<z<2$ for our half-mass radii argue in favor of our conclusion that the sizes of massive quiescent galaxies strongly evolve with cosmic time, in line with previous results based on rest-frame optical size measurements \citep[e.g.,][]{trujillo04, trujillo06b, van-der-wel08b, van-dokkum08a, newman12, carollo13, van-der-wel14}.

\section{Summary \& Outlook}
Our novel method to estimate stellar half-mass radii for a large sample of galaxies at $0.5<z<2.5$ drawn from CANDELS and 3D-HST rests on leveraging the integrated UV-to-midIR photometric information that is available for these galaxies. We derive a relationship between the HST/ACS$+$WFC3 colors and the $M/L$ estimated from the UV-to-midIR SED (Sec.~\ref{sec:colorml}) and apply that relationship to the spatially resolved color profiles (Sec.~\ref{sec:conversion}). The underlying assumption is that the distribution of physical properties (age, metallicity, attenuation, etc.) among galaxies is comparable to that within galaxies. Moreover, we infer 3D sizes based on the deprojection machinery developed by \citet{van-de-ven21} and our knowledge of the shape distribution of galaxies and its dependence on stellar mass and redshift (Sec.~\ref{sec:deproject}).  The \rmass~and \rmassd~estimates are made publicly available online -- see Table \ref{tab:cat1} for the first 10 entries of the catalog. 

An essential test of the reliability of our stellar half-mass radii is provided by the comparison with size measurements from JWST/NIRCam imaging in the rest-frame near-IR for a (for now) limited subset galaxies in CEERS. The agreement is excellent (Sec.~\ref{sec:nircam}). First, systematic offsets are less than 10\% for quiescent galaxies up to $z=2$ and for star-forming galaxies up to $z=1.5$. Second, the scatter is small (typically, 25\%) and consistent with the formal error budget. The comparison with NIRCam demonstrates that our stellar half-mass radii are precise and accurate under the assumption that rest-frame near-IR sizes are, indeed, a good proxy for stellar-mass weighted sizes. As briefly discussed in Section \ref{sec:nircam} this is not self-evident. Even in the near-IR the $M_\star/L$ evolves strongly with age. Either our \rmass and $R_{\rm{NIR}}$ sizes agree because they both accurately trace the stellar mass distribution, or they suffer from the same systematic bias. The latter is a distinct possibility: if our $M_\star/L$ are overestimated in regions with high star-formation activity (see Sec.~\ref{sec:colorml}) then lower $M_\star/L_{\rm{NIR}}$ are to be expected as well.  

Previously published half-mass radii do not perform equally well, as described in Sec.~\ref{sec:evol}, with larger systematic offsets, and larger scatter while reporting smaller formal uncertainties. The main caveat of the present analysis is that for small objects $\lesssim$~1 kpc the systematic uncertainties are not well understood, not only because this is near the resolution limit of HST in the optical and JWST/NIRCam in the near-IR, but also because the NIRCam PSF is not sufficiently well understood at the moment.

In Section \ref{sec:mr} we show the effects of correcting for $M_\star/L$ gradients and deprojecting the 2D distribution on the size-mass distribution of galaxies at $0.5<z<1.5$. Compared to the rest-frame optical size distribution, the stellar half-mass radius - stellar mass relation is less steep (Fig.~\ref{fig:m_r}), while deprojection affects the size-mass distribution only little. A separation between star-forming and quiescent galaxies (Sec.~\ref{sec:sfq}) shows that the flattening of the size-mass relation is driven by massive star-forming galaxies, which have the largest downward $M_\star/L$ correction.

For quiescent galaxies, the deprojection counters the downward $M_\star/L$ and the size-mass distribution in the rest-frame optical is very similar to the 3D half-mass radius distribution, modulo a small ($\approx 0.05$~dex) downward shift.  The medians and percentile values of the various size distributions shown in Figures \ref{fig:m_r}, \ref{fig:m_r_sf} and \ref{fig:m_r_q} are provided in Tables \ref{tab:medians}, \ref{tab:medians_sf}, and \ref{tab:medians_q}, respectively. In Section \ref{sec:evol} we show that the average \rmass and \rmassd of massive, quiescent galaxies evolve rapidly from $z=2.3$ to $z=0.5$, with $R\propto (1+z)^{-1.7\pm0.1}$, and at the same pace as the rest-frame optical half-light radii.

Now that NIRCam imaging datasets across larger volumes become available from COSMOS-Web \citep{casey22} and JADES \citep{robertson22}, and the sample sizes become similar to those drawn from CANDELS, then our stellar mass profiles will be superseded by rest-frame near-IR profiles and the modeling of the optical-to-near-IR light profiles as pioneered by \citet{miller22, abdurrouf23, ji23}. But our work provides a simple conversion from light-to-stellar mass weighted sizes for galaxies without spatially resolved near-IR imaging, that is, those without NIRCam imaging or those at high redshift ($z>6$) when even NIRCam only samples the rest-frame optical. The key point of our work is that spatially resolved optical colors accurately predict the sizes of galaxies in the rest-frame near-IR, which is generally considered a robust proxy for the stellar mass distribution.

\section*{Acknowledgments}
MM acknowledge the financial support of the Flemish Fund for Scientific Research (FWO-Vlaanderen), research project G030319N. 
All the {\it HST} and {\it JWST} data used in this paper can be found in MAST: \dataset[10.17909/z7p0-8481]{http://dx.doi.org/10.17909/z7p0-8481}, \dataset[10.17909/T94S3X]{http://dx.doi.org/10.17909/T94S3X}.

\bibliographystyle{aasjournal}
\bibliography{ms}

\appendix
\section{Comparison with \rmass~Estimates from the Literature}
\restartappendixnumbering

\begin{figure*}[!t]
\epsscale{1.15}
\plotone{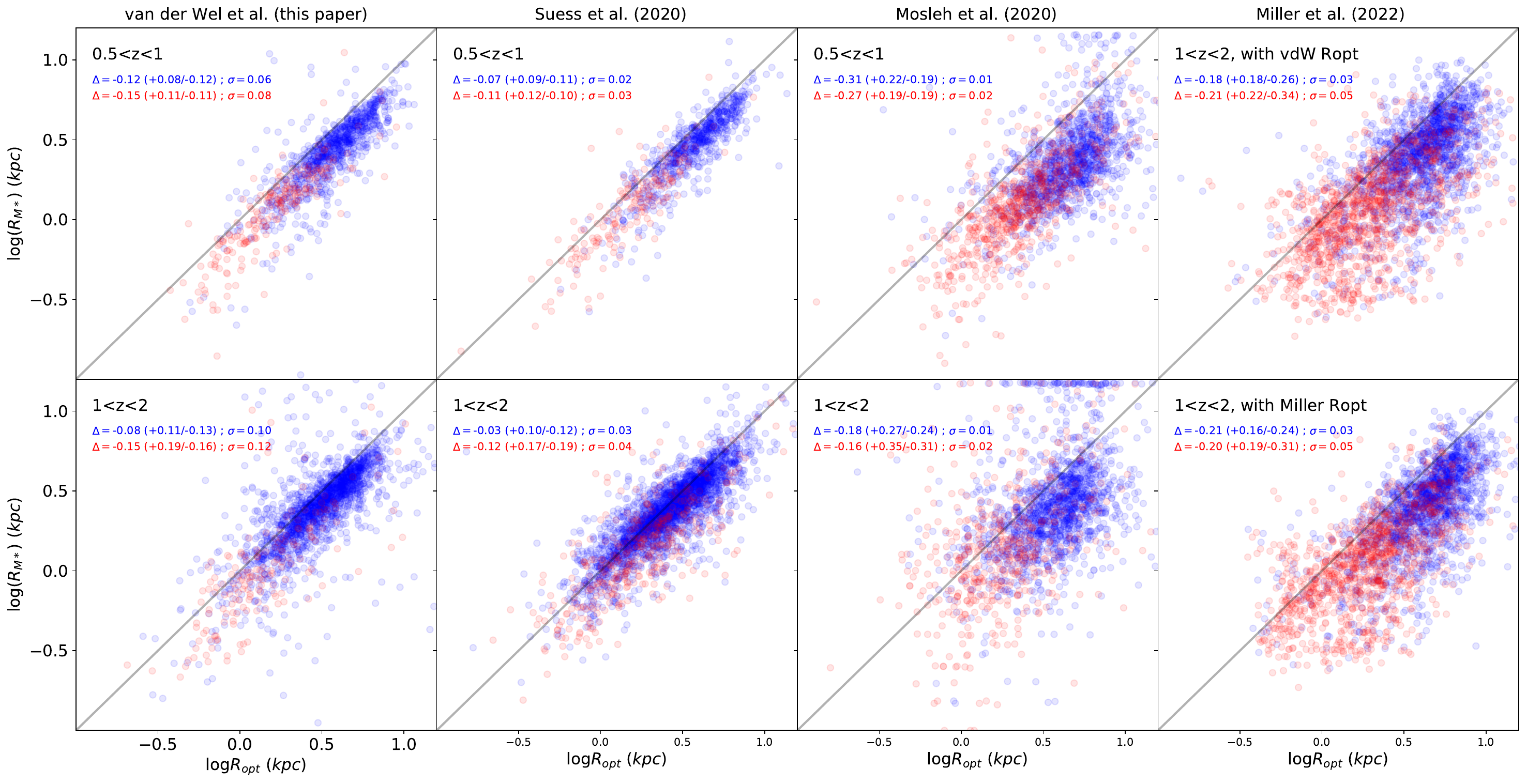}
\caption{Comparison of rest-frame optical half-light sizes and stellar-half mass radii, equivalent to Figure \ref{fig:rm_ropt}, but now for  four different estimates of the half-mass radii as labeled above the panels. \ropt~is taken from the respective authors. Bottom panels show results for $1<z<2$; top panels show results for $0.5<z<1$, with the exception of the top-right panel: \citet{miller23} did not provide \rmass~estimates for $z<1$. Instead, here we show the comparison of the \citet{miller23} \rmass~estimates and the \ropt~estimates used in this paper (the bottom panel compares with the \citet{miller23}~\ropt estimates). Results are shown for quiescent (red) and star-forming (blue) galaxies with total stellar mass $>10^{10}~M_\odot$. $\Delta$ is the median offset in dex, with the 16-84\%-ile scatter in parentheses. $\sigma$ is the median formal uncertainty as reported by the respective authors.
\label{fig:rm_ropt_all}}
\end{figure*}

\begin{figure*}[!t]
\epsscale{1.15}
\plotone{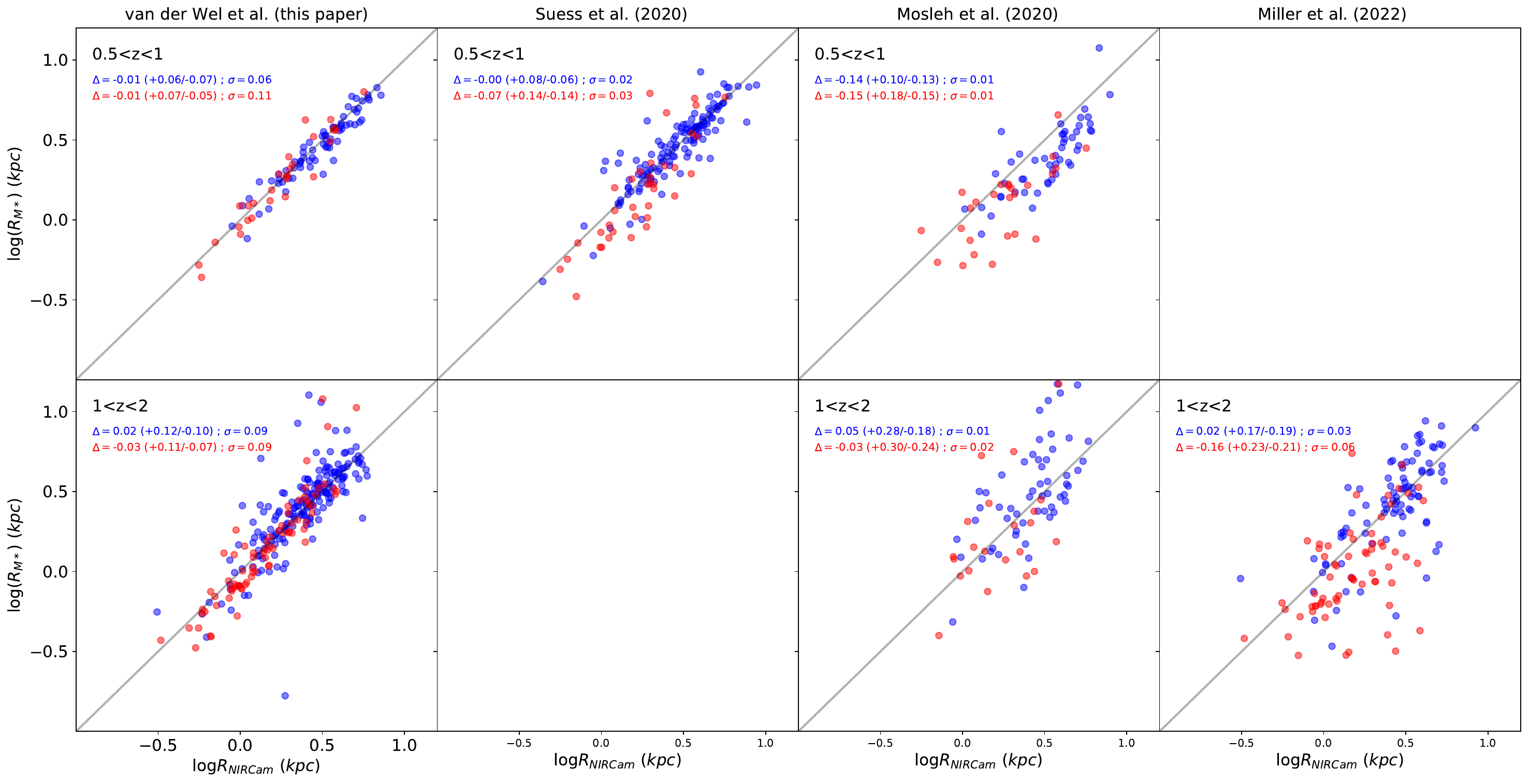}
\caption{Comparison of rest-frame near-infrared half-light sizes from NIRCam and stellar-half mass radii, equivalent to Figure \ref{fig:rm_ropt}, but now for  four different estimates of the half-mass radii as labeled above the panels. Results are shown for quiescent (red) and star-forming (blue) galaxies with total stellar mass $>10^{10}~M_\odot$. $\Delta$ is the median offset in dex, with the 16-84\%-ile scatter in parentheses. $\sigma$ is the median formal uncertainty as reported by the respective authors. \citet{suess20} does not provide $z>1$~\rmass~estimates for galaxies in the CEERS footprint; \citet{miller23} does not provide $z<1$~\rmass~estimates.
\label{fig:rm_rnir_all}}
\end{figure*}

Previously published estimates of stellar half-mass radii based on CANDELS data are shown in Figures \ref{fig:rm_ropt_all} and \ref{fig:rm_rnir_all} in the same manner as in Figures \ref{fig:rm_ropt} and \ref{fig:rm_rnir}. 

Figure \ref{fig:rm_ropt_all} compares, for four different authors, the rest-frame optical half-light radii $R_{\rm{opt}}$\footnote{Except for our own estimates, which are at rest-frame 0.5$\mu$m, the plotted values are measured from HST/WFC3/F160W CANDELS imaging} with half-mass radii \rmass, each time comparing the radii as published by the authors. The one exception is the top-right panel (see figure caption for details). Even though all \rmass~estimates are systematically smaller than $R_{\rm{opt}}$, the offsets and scatter vary from author to author. Our estimates show similar offsets and scatters as those by \citet{suess20}, adopting their preferred Method 1 estimates. The main difference is that their uncertainties are several times smaller than ours. The \rmass~estimates by \citet{mosleh20} and \citet{miller23} show much larger scatters, and very small formal uncertainties ($\lesssim 0.05$~dex). Taken at face value this means that these authors find a large (0.2$-$0.3 dex) galaxy-to-galaxy scatter in the \rmass$/$\ropt~(in fact, larger than the scatter in $R_{\rm{opt}}$ at fixed mass). 

We note that in all cases the $R_{\rm{opt}}$ agree well between the authors, with small scatter and no systematic offsets. The largest scatter (0.1 dex) is found when comparing with \citet{miller23}, who model the light profiles in a fundamentally different manner (multi-gauss expansion rather than S\'ersic profile fitting). These differences do not explain the different trends and patterns seen in Figure  \ref{fig:rm_ropt_all} -- rather, those are due to the variety in techniques to correct for $M_\star/L$~gradients (see discussion in Sec.~\ref{sec:methods}).

The comparison with NIRCam-based rest-frame near-infrared sizes used in this paper provides additional insights, as illustrated in Figure \ref{fig:rm_rnir_all}. As already demonstrated in Section \ref{sec:nircam} the \rmass~estimates presented in this paper compare well with the NIRCam sizes, with a reasonably small scatter of similar magnitude as the formal uncertainties. The \rmass~estimates by \citet{mosleh20} and \citet{miller22} show much larger scatter, especially for quiescent galaxies (0.2-0.3 dex) suggesting low precision, especially when compared to the very small formal uncertainties (0.01-0.06 dex).  

Unfortunately, \citet{suess20} \rmass~estimates for the CEERS NIRCam sample are not available for $z>1$ and the comparison is limited to $z<1$. For those galaxies the \citet{suess20}~\rmass~estimates agree fairly well with the NIRCam sizes, with somewhat larger scatter than our \rmass estimates. It should be kept in mind that light profiles used by \citet{suess20} are not identical to our profiles so that the comparison is not entirely fair and straightforward. The scatter in \ropt between \citet{suess20} and ours is $0.05$~dex, which can explain the difference in scatter for the star-forming galaxies. The sample of quiescent galaxies is too small to make a reliable statement.

We must now address the question where the tension arises between the conclusions presented by \citet{suess20}, who find little or no evolution in \rmass for quiescent galaxies at $z>1$ and the results presented here in Figure \ref{fig:rz_massive}, with strong evolution in \rmass for massive quiescent galaxies up to $z=2.3$. The size comparisons in Figures \ref{fig:rm_ropt_all} and \ref{fig:rm_rnir_all} do not provide immediate answers.

\begin{figure*}[!t]
\epsscale{0.55}
\plotone{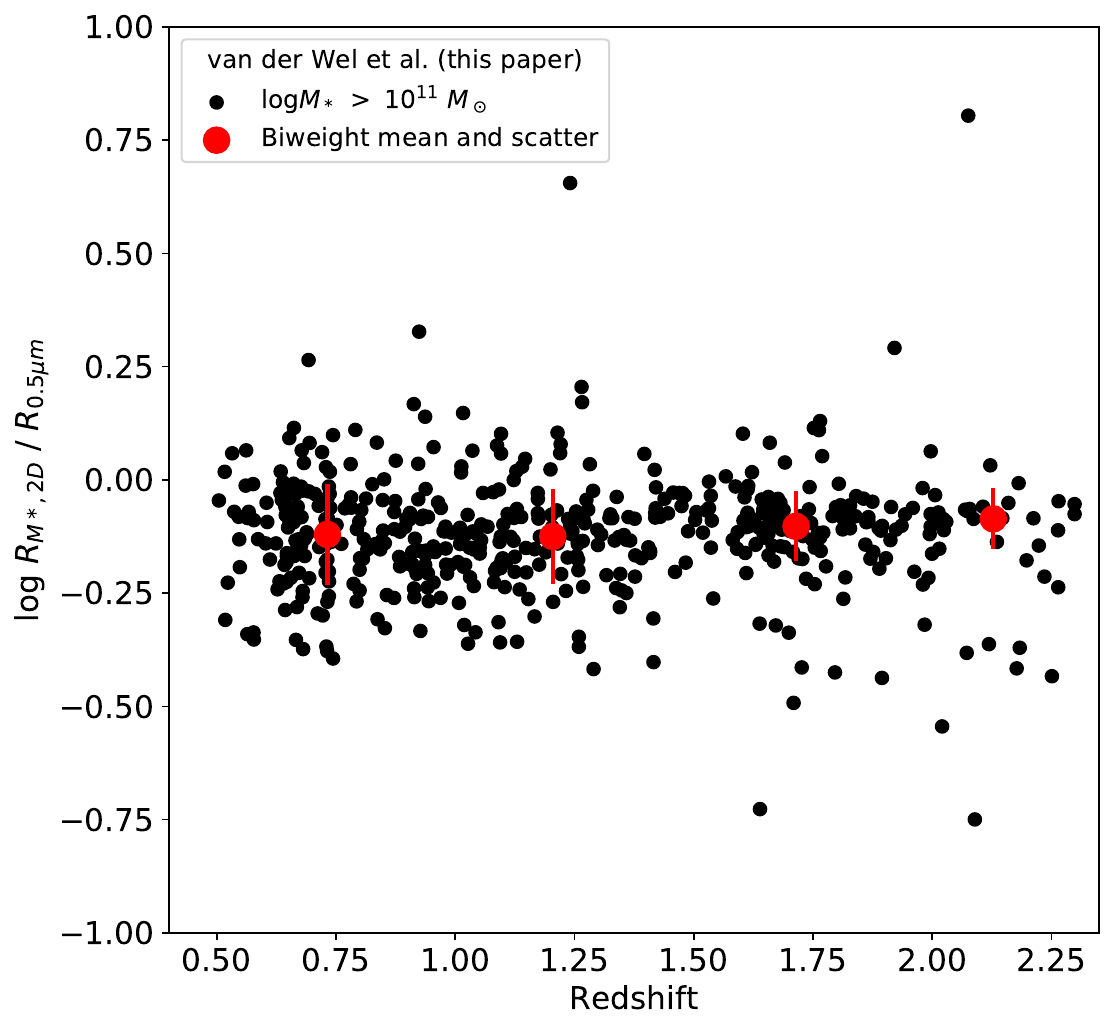}
\plotone{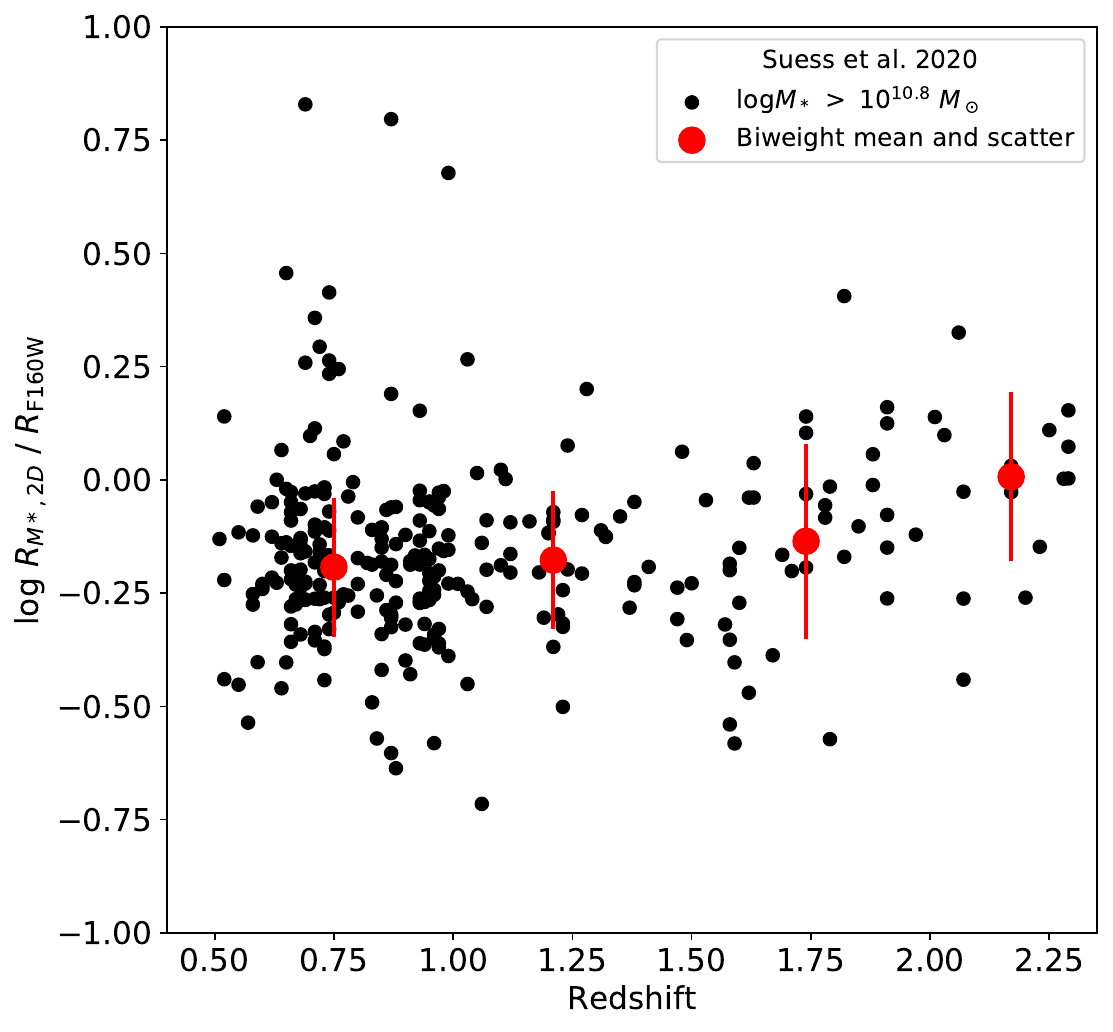}
\caption{Redshift evolution of the ratio \rmass$/$\ropt~of massive quiescent galaxies. \textit{Left:} Results from this paper; \textit{Right:} Results from \citet{suess20}. Due to the systematic difference in $M_\star$ the stellar mass limit is shifted accordingly (see text for discussion). 
\label{fig:rz_massive_ratio}}
\end{figure*}

Figure \ref{fig:rz_massive_ratio} shows the evolution in the light-to-mass weighted size ratio (\rmass$/R_{\rm{opt}}$), comparing the results presented in this paper and those from \citet{suess20}. In both cases the measurements from the respective authors are used without matching catalogs. Differences in redshift measurements, stellar mass estimates and size estimates can all contribute to differences in the comparison. Notably, the stellar mass estimates used here are systematically larger by 0.2 dex, which is here accounted for by lowering the stellar mass cut. Up to $z\approx 2$ the patterns are similar, with \rmass~estimates that are 0.1-0.2 dex lower than \ropt, both works showing no evidence for a different pace of evolution in \rmass~compared to \ropt. 

The main difference arises at $z>2$, where \citet{suess20} find no offset and we do. For \citet{suess20} the scatter in \rmass$/$\ropt at $z>1.5$ ($>$0.2 dex)is larger than the scatter in $R_{\rm{opt}}$. For our own estimates this is not the case, but we see an increased number of outliers at $z>1.5$. These trends suggest that uncertainties start dominating over the corrective effect of accounting for $M_\star/L$~gradients at $z>2$. A more definitive statement on the evolution of \rmass~beyond $z=2$ will have to wait for NIRCam size measurements for larger samples.

\renewcommand{\thefigure}{\arabic{figure}}

\end{document}